\documentclass[prb,english,amsmath,notitlepage,twocolumn,showpacs,superscriptaddress,floatfix]{revtex4-1}

\usepackage[pdftex]{graphicx}
\usepackage[usenames,dvipsnames]{color}
\usepackage{epstopdf}
\usepackage{longtable}
\usepackage{widetable}
\usepackage{amssymb,braket}
\usepackage{ulem}

\definecolor{green}{rgb}{0,.5,0}

\usepackage[pdftex,bookmarks,breaklinks,bookmarksopen,bookmarksnumbered,colorlinks,linkcolor=purple,linktocpage,citecolor=blue,urlcolor=blue,pdfpagemode=UseOutline,pdftex]{hyperref}

\usepackage[colorinlistoftodos,prependcaption,textwidth=1cm,textsize=footnotesize,color=lime]{todonotes}

\newcommand{\cDW}{cDW{}}
\newcommand{\dDW}{dDW{}}
\newcommand{\Imag}{\mathcal{I}m{}}

\definecolor{darkblue}{rgb}{0.0, 0.0, 0.55}

\begin{document}

\title{Classification of magnetic inhomogeneities and $0-\pi$ transitions\\ in superconducting-magnetic hybrid structures}
\author{Thomas E. Baker}
\affiliation{Department of Physics \& Astronomy, University of California, Irvine, CA 92697}
\author{Adam Richie-Halford}
\affiliation{Department of Physics \& Astronomy, University of Washington, WA 98195}
\author{Andreas Bill}
\email[Author to whom correspondence should be addressed: ]{abill@csulb.edu}
\affiliation{Department of Physics \& Astronomy, California State University Long Beach, CA 90840}
\date{June 15, 2016}

\begin{abstract}
We present a comparative study of pair correlations and currents through superconducting-magnetic hybrid systems with a particular emphasis on the tunable Bloch domain wall of an exchange spring. This study of the Gor'kov functions contrasts magnetic systems with domain walls that change at discrete points in the magnetic region with those that change continuously throughout. We present results for misaligned homogeneous magnetic multilayers, including spin valves, for discrete domain walls, as well as exchange springs and helical domain walls --such as Holmium-- for the continuous case.  Introducing a rotating basis to disentangle the role of singlet and triplet correlations, we demonstrate that substantial amounts of (so-called short range) singlet correlations are generated throughout the magnetic system in a continuous domain wall via the cascade effect. We propose a classification of $0-\pi$ transitions of the Josephson current into three types, according to the predominant pair correlations symmetries involved in the current. Properties of exchange springs for an experimental study of the proposed effects are discussed. The interplay between components of the Gor'kov function that are parallel and perpendicular to the local magnetization lead to a novel prediction about their role in a proximity system with a progressively twisting helix that is experimentally measurable.
\end{abstract}


\maketitle

\section{Introduction}\label{s:intro}

There has been an accrued interest in nanoscale proximity systems made of materials with competing ground states, and in particular those involving  superconductivity and magnetism. The interest is both fundamental and practical: The behavior of conduction electrons in competing phases of matter provides fertile ground for rich physics\cite{buzdinRMP05,bergeretRMP05} and are also candidates for new spintronics devices.\cite{eschrigPT11,linderNP15,eschrigRPP15,gingrichNP16,martinezPRL16}

The interplay of superconductivity and magnetism is an old topic. The first result related to the present work is the Fulde, Ferrell, Larkin, Ovchinnikov (FFLO) effect in which Cooper pairs entering a magnetic field or material will acquire an angular momentum and change the state of the spin pairs.\cite{fuldePRL64,larkinJETP65} Denoting $\ket{s,m}$ the spin state of the pair with $s=0,1$ the total spin and $m=0,\pm 1$ the projection of that spin onto the quantization axis, the FFLO effect transforms the singlet $\ket{0,0}$ state into a linear combination of both $m=0$ states, $\ket{0,0}$ and $\ket{1,0}$.
Of particular interest for technological application is the generation of triplet pairs $\ket{1,\pm 1}$ that have a long propagation length scale.  These pair correlations appear in the presence of magnetic inhomogeneities. 
This was pointed out by Bergeret, Volkov, Efetov and Kadigrobov, Shektar, Jonson.\cite{bergeretPRL01,kadigrobovEPL01}  Experimental verification of the theoretical predictions showed that they play a crucial role in the detection of superconducting properties in wide ferromagnets.\cite{anwarPRB04,keizerN06,khairePRL10,robinsonS10,zhuPRL10,leksinPRL12,klosePRL12,wenEPL14,khasawnehSST11}

We discuss fundamental properties of pair correlations in superconducting-magnetic hybrid systems in the diffusive regime. Of interest is the behavior of spin pair-correlations in various inhomogeneous magnetic heterostructures. The particularity of these systems is that the quantization axis changes direction in the structure, thereby affecting pair correlations at the nanoscale.

Prior theoretical works have discussed proximity effects in misaligned homogeneous films \cite{bergeretPRL01,bergeretRMP05,buzdinRMP05,houzetPRB05,houzetPRB07,haltermanPRL07,haltermanPRB08,bakerAIP16,bakerEPL14} or rotating magnetizations \cite{bergeretPRL01,bergeretPRB01,bergeretRMP05,buzdinRMP05,cretinonPRB05,eschrigJLTP07,linderPRB09,alidoustPRB10,alidoustPRB10b,richardPRL13,zhuPRL13,wuPRL12,wuPRB12,fritschNJP14,fritschJPC14,bakerNJP14} in the clean or the diffusive limit.  The results strongly depend on the choice of materials and the width of the layers.

On the experimental side, mostly misaligned homogeneous films have been studied.\cite{khairePRL10,khasawnehSST11}
Some also considered conical magnetization profiles such as those encountered in Holmium (Ho) or an exchange spring.\cite{robinsonS10,zhuPRL13,guPRB10,linderPRB09,alidoustPRB10,alidoustPRB10b,fritschNJP14,fritschJPC14}

This work provides a comparative study of pair correlations in existing and proposed hybrid systems, including spin valves and other multilayers of homogeneous Fs, helical magnetic structures and exchange springs (XS). Particular focus is set on the latter that was proposed as a device to tune and reverse the Josephson current in one single heterostructure.\cite{bakerNJP14} The insight provided by the study of pair correlations allows understanding which linear combination of spin states is dominant and under what circumstances. To this aim, we focus on a discussion of the Gor'kov functions that describe the superconducting pair correlations.  These functions are not usually presented in the literature, yet they provide for a clear picture of the behavior of electron pairs in proximity systems and Josephson junctions. We introduce a rotating basis that follows the magnetization direction to disentangle the behavior of either singlet and triplets, or $m=0$ and $m\neq 0$ pair correlations at each point in the magnetic structure.

The study of pair correlations in various structures leads to two main insights. First, we are led to divide magnetic inhomogeneities into discrete domain walls (\dDW) and continuous domain walls (\cDW), which refer to local abrupt and continuous, smooth rotations of the magnetization, respectively.
Second, we classify all types of $0-\pi$ transitions of the Josephson current (the reversal of the current upon variation of a parameter of the system)  according to the symmetry of the pairs correlations involved in the current (summarized in table \ref{tab:0pi} of Sec.~\ref{ss:classificationcurrent})

The distinction between \dDW and \cDW s is rooted in the fact that pair correlations propagate in fundamentally different ways through these magnetic inhomogeneities. The most dramatic effect is seen on the singlet component. The singlet correlations are found to be present throughout the magnetic material of a \cDW\ despite their known short decay length. This is due to the cascade effect, which is a remix of all pair correlations when the magnetization changes direction.\cite{bakerEPL14} Also, despite their scalar (rotationally invariant) nature, singlet correlations are shown to be affected by the magnetic configuration of the hybrid structure and not only by the magnitude of the magnetization in cDWs.

The symmetry of pair correlations is also determinant for understanding how the $0-\pi$ transition of the Josephson current comes about. We distinguish three classes of Josephon current reversal. One relies on the $m=0$ components and was proposed by Buzdin, Bulaevskii and Panyukov.\cite{buzdinJETP82} Another class of $0-\pi$ transition involves  only $m\neq0$ triplet components and was presented by Houzet and Buzdin.\cite{houzetPRB07} Finally, the third class of $0-\pi$ transitions involves a mixture and competition of singlet and triplet correlations and was discussed in Ref.~\onlinecite{bakerNJP14}. Earlier indication of the existence of the third class can be found in the work of Bergeret, Volkov and Efetov in Ref.~\onlinecite{bergeretPRB01}, although the transition of that paper is a blend of the Buzdin-Bulaevskii-Panyukov and the singlet-triplet $0-\pi$ transition discussed here.

\begin{figure}
\includegraphics[width=\columnwidth]{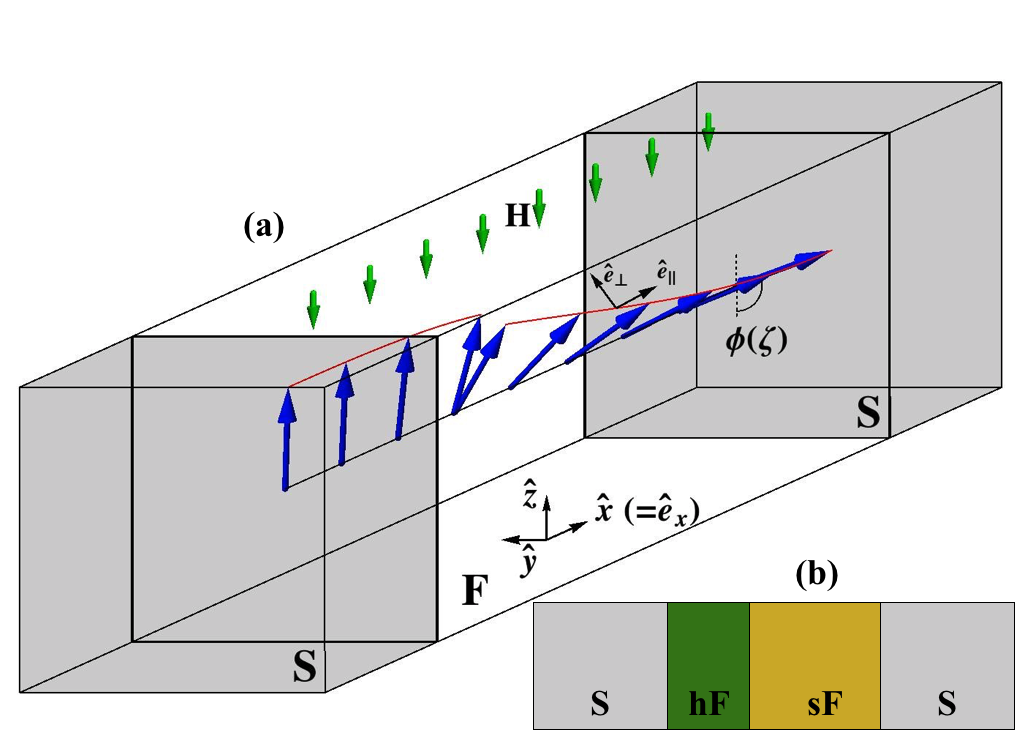}
\caption{\label{fig:XS} (color online) (a) A perspective of the S/XS/S hybrid system.  The exchange spring (XS) is made of two Fs with easy axes parallel to $\hat{\mathbf{z}}$. The hard F (hF) has a high anisotropy energy $K_h$ as compared to the soft F (sF) energy $K_s$.\cite{chikazumi1978physics} (b) A schematic side representation of the S/XS/S system. Shown in (a) are also the Cartesian coordinate system $\{\hat{\mathbf{x}},\hat{\mathbf{y}},\hat{\mathbf{z}}\}$, and the rotating basis $\{\hat{\mathbf{e}}_x\equiv \hat{\mathbf{x}},\hat{\mathbf{e}}_\perp(x),\hat{\mathbf{e}}_\parallel (x) \}$ where $\perp$ ($\parallel$) denote the vectors perpendicular (parallel) to the local magnetization vector $\mathbf{h}(x)$.}
\end{figure}

Using our own numerical approach to solve the complete, non-linear Usadel equation for the wide limit in the diffusive regime (described in Sec.~\ref{sec:model}), we determine the pair correlations and discuss the physics of various observed and predicted effects on hand of the exchange spring (XS; Refs.~\onlinecite{billJMMM04,bakerNJP14}) pictured in Fig.~\ref{fig:XS}.
This magnetic bilayer proposed earlier by the authors to vary the Josephson current\cite{bakerNJP14} has the advantage of being magnetically tunable, allowing to change the relative weight of the pair correlations within the same system. The other advantage is that it is experimentally realizable and has been studied at length in the field of magnetism.\cite{billJMMM04,guPRB10} After describing the pair correlations in the XS (Sec.~\ref{sec: SXSS}), we compare in Sec.~\ref{sec:hybridsystems} the XS and other structures studied in the literature, such as helical $\cos(Qx)$ type domain walls and misaligned homogeneous multilayers F$_1$F$_2\cdots$.  For concreteness, we compare the pair correlations of our XS with that of the helical structure of Refs.~\onlinecite{bergeretPRB01,wuPRL12,wuPRB12,fritschNJP14,fritschJPC14}, and misaligned multilayers of Ref.~\onlinecite{houzetPRB07}. We set particular emphasis on identifying which pair correlations drive the behavior of the proximity system.

In Secs.~\ref{ss:generalJosephson}-\ref{ss:singlet-triplet} we consider how the behavior of the Gor'kov functions in \cDW\ and \dDW\ hybrid structures differs and affects the Josephson current. Table \ref{tab:0pi} summarizes the results of this study and unambiguously differentiates the three classes of transitions.

We discuss in Sec.~\ref{sec: expt} materials properties that affect the measurement of the Josephson current through an XS and propose alternative experiments that discuss new effects in superconducting-magnetic hybrid structures. We conclude in Sec.~\ref{sec:conclusions}.

\section{Superconducting Proximity Effects: The Model}\label{sec:model}

We consider proximity effects in which a singlet-pair superconductor is in contact with a magnetic material. The Cooper pairs from the superconductor may tunnel into the adjacent material, but the absence of a pairing mechanism will cause the probability amplitude to attenuate with distance. In the diffusive regime the length scale over which the exponential decay occurs is the coherence length $\xi_N=\sqrt{D_N/2\pi T}$ for non-magnetic metals ($D_N$ is the diffusion constant and $T$ the temperature) and $\xi_F=\sqrt{D_F/h}$ for ferromagnetic materials ($h$ is the magnetization of the ferromagnet). Another coherence length, $\xi_c=\sqrt{D_F/2\pi T_c}$ where $T_c$ is the superconducting critical temperature of the proximity system, is also introduced as a length scale available when analyzing two ferromagnets of different strengths.\cite{billJSNM12,buzdinRMP05}

\subsection{Pair correlations and Josephson current}

The standard Fermi-surface and impurity averaged Green, $g$, and Gor'kov, $f$, functions are used to analyze the state of the system.\cite{AGD,BCS,gorkovJETP59,eilenbergerZP68,chandrasekhar2008proximity,bennemann2008superconductivity,kopnin2009theory,belzigS&M99,demlerPRB97,bergeretRMP05,buzdinRMP05,golubovRMP04,deutscher1969superconductivity,ARHMaster,bakerMaster}  These satisfy the Usadel equations in the semi-classical diffusive regime,\cite{usadelPRL70} where the elastic scattering length is much smaller than the coherence length of the superconducting pairs, effectively randomizing the momentum of the electrons.

To describe the possible correlations of the pairs of spin$-1/2$ particles one writes the Gor'kov function, $f$, in the L\"uder's expansion to sufficient order $f =f_0+\mathbf{\hat v}\cdot\mathbf{f}+\ldots$\cite{luders1971method} where $\mathbf{\hat v}$ is the unit-vector along the Fermi velocity.\cite{usadelPRL70}  The scalar term, $f_0$, correponds to singlet pair correlations while the vector function $\mathbf{f}$ describes triplet states.  A similar decomposition is performed for the Green's function, $g$.\cite{buzdinRMP05,bergeretRMP05}

We are interested in the behavior of pair correlations in magnetic systems where the magnetization $\mathbf{h}$ is confined to the $yz$ plane (the plane of the thin films) and may rotate in that plane as a function of $x$ (see Fig.~\ref{fig:XS}). The rotationally invariant singlet component, $f_0$, is in principle only affected by the magnitude $h$ of the magnetization and not the direction, while the vector function $\mathbf{f}$ is affected by both the magnitude and direction of $\mathbf{h}$.
The equations governing the Gor'kov functions in a magnetic system (see Sec.~\ref{sec:UsadelIN}) reveal that components of $\mathbf{f}$ parallel to $\mathbf{h}$ are affected by the magnetization while the components perpendicular to it remain unaffected.\cite{ivanovPRB06} Since we consider magnetic configurations in which the magnetization vector $\mathbf{h}(x)$ rotates in the $yz$ plane, we present the Gor'kov vector $\mathbf{f}(x)$ either in the fixed Cartesian system or in the rotating basis $\{\hat{\mathbf{x}},\mathbf{e}_\perp(x),\mathbf{e}_\parallel(x)\}$ (see Fig.~\ref{fig:XS}), where the $\perp$ ($\parallel$) index denotes the component perpendicular (parallel) to the local direction of the magnetization $\mathbf{h}(x)$ in the $yz$ plane. Thus, we write the vector Gor'kov function $\mathbf{f} = f_y(x) \, \hat{\mathbf{y}} +  f_z(x)\,\hat{\mathbf{z}}$ in the standard fixed Cartesian coordinate system $\{\hat{\mathbf{x}}, \hat{\mathbf{y}}, \hat{\mathbf{z}}\}$.
Introducing the angles $\gamma(x) = \angle(\mathbf{h},\mathbf{f})$ and $\phi(x) = \angle(-\mathbf{\hat{z}},\mathbf{h})$ we have 
\begin{eqnarray}
\mathbf{f}(x)
&=& |\mathbf{f}| \left( 0, \sin(\phi-\gamma), \cos(\phi-\gamma) \right)_{xyz}\\
&=& |\mathbf{f}| \left( 0, \sin\gamma, \cos\gamma \right)_{x,\perp,\parallel}
\end{eqnarray}
 in the Cartesian and rotating basis, respectively.
With the magnetization confined to the $yz$ plane we have
\begin{eqnarray}\label{rotatingbasis}
\left(
 \begin{array}{c}
 f_\perp(x)\\
 f_\parallel(x)
 \end{array}
\right)
=
-
\left(
\begin{array}{cc}
\cos\phi(x) & \sin\phi(x) \\
-\sin\phi(x) & \cos\phi(x)
\end{array}
\right)
\left(
 \begin{array}{c}
 f_y(x)\\
 f_z(x)
 \end{array}
\right).
\end{eqnarray}

All calculations presented in this paper are performed in the wide limit where effects within one coherence length are smeared out.\cite{zaikinZP81,belzigS&M99,vasenkoPRB08} This is a reasonable limit for our comparative study of XSs and related systems because they are tens of nanometers wide and the physics of interest lies far from the SF interface. If one desires to remove this approximation, a position dependent spinor must be included in the Gor'kov function and different boundary conditions enforced.\cite{zaikinZP81,belzigS&M99,houzetPRB05,vasenkoPRB08}
An example effect that is excluded  in this approach is the inverse FFLO (or inverse proximity) effect, where pairing in the S region is reduced by the magnetization in the adjacent F.\cite{golubovRMP04,lofwanderPRB07}

The wide limit treatment of Usadel's equation has the advantage that it allows for a clear determination of what is contributed from the left (L) and right (R) superconductors, respectively, in a Josephson junction. Following Refs.~\onlinecite{zaikinZP81} and \onlinecite{vasenkoPRB08}, we may thus write ($\alpha=0,x,y,z$)
\begin{equation}\label{leftright}
f_\alpha(x)=e^{i\varphi/2}f_{\alpha,L} + e^{-i\varphi/2} f_{\alpha,R}.
\end{equation}
where $\varphi = \varphi_L - \varphi_R$ is the superconducting phase difference.

In this work, we typically show the Gor'kov functions for the first Mastubara frequency. This is beneficial for studying pair correlations across the system.  Higher Matsubara frequencies see an overall decrease in the superconducting order parameter amplitude which results in a  decrease of the Gor'kov function while slightly shifting some of its features ({\it e.g.} dips, zeros, etc.) in space.  When calculating the Josephson current, one sums over all frequencies as the features of the Gor'kov functions for all frequencies count. Nevertheless, the Gor'kov functions for the first Matsubara frequency can reveal much about the electronic state of the system.

Once the Gor'kov functions have been calculated. We can determine the Josephson critical current of a junction. All effects discussed in this paper are found in the first harmonic. The superconducting critical current is given by\cite{buzdinRMP05,houzetPRB07,champelPRL08}
\begin{eqnarray}\label{IcRNtot}
I_c(x) = \frac{ \pi T}{2e R_N} \sum_{n = -\infty}^\infty \Imag\left( f_{-n}^\star\frac{\partial f_{n}}{\partial x}\right).
\end{eqnarray}
This expression may be rewritten as\cite{jcleftrightnote}
\begin{eqnarray}\label{IcRNtotdecomp}
I_c(x)&=&\frac{\pi T}{2eR_N}\sum_{n=-\infty}^\infty\sum_{\alpha=0,x,y,z}\Imag\left(f_{-n,\alpha}^*\frac{\partial f_{n,\alpha}}{\partial x}\right)\nonumber\\
&=&\left[I_{c,0}(x)+I_{c,t}(x)\right]\sin\varphi.
\end{eqnarray}
The current is sinusoidal in $\varphi$ having neglected the inverse FFLO effect due to opaque boundary conditions,\cite{vasenkoPRB08}  though others may be chosen.\cite{faurePRB06}

In the first line we have expressed $I_c$ in terms of a sum over each component $0,x,y,z$ of the L\"uders decomposition. The second line shows that the same current can be decomposed in contributions from the singlet, $I_{c,0}$, and from all triplets, $I_{c,t}=\sum_{\alpha=x,y,z}I_{c,\alpha}$. 
For magnetic configurations confined to the $yz$ plane we also have $I_{c,t} = \sum_{\alpha = \perp,\parallel} I_{c,\alpha}$.

\subsection{Solution for the homogeneous magnetization}\label{sec:usadelhom}

We point out a few particularities of the well known homogeneous case to set the stage of the theory and for later comparison.
The standard parametrization of the Green and Gor'kov functions in a homogeneous F, $g = \cos \theta$, $f = \sin \theta$ with $\theta = \theta_R + i \theta_I$, guarantees that $\hat g = g \hat\sigma_z + f \hat\sigma_x$ (with Pauli matrices $\hat \sigma$ for the spin sector) automatically satisfies the normalization condition, $\hat g^2=\hat{1}$.\cite{langenbergBook86,buzdinRMP05,shelankovJLTP85}
Usadel's equation in the wide limit and for homogeneous magnetization $\mathbf{h} = h\hat{\mathbf{z}}$ then takes the form \cite{zaikinZP81,vasenkoPRB08,bakerEPL14,faurePRB06,houzetPRB05}
\begin{equation}\label{usadelhom}
D\partial_x^2\theta=2(\beta+\cos\theta/\tau)\sin\theta
\end{equation}
where $\beta=\omega_n+i\,\mathrm{sgn}(h)$, $D$ is the diffusion coefficient of the medium (subscripted $F$ for ferromagnet, $S$ for superconductor) and $\tau$ is the spin--flip scattering time if magnetic impurities are present.  The $n^\mathrm{th}$ fermionic Matsubara frequency at temperature $T$ is $\omega_n=(2n+1) \pi T$.  Equation \eqref{usadelhom} was solved exactly analytically for arbitrary values of the parameters and finite thicknesses in closed form in Ref.~\onlinecite{bakerEPL14}, and Refs.~\onlinecite{cretinonPRB05,faurePRB06,oboznovPRL06}
or by linearizing the equation near the critical temperature.\cite{zaikinZP81,vasenkoPRB08,faurePRB06,houzetPRB05}  Some expressions have been shown to fit experimental data.\cite{oboznovPRL06}

For a normal metal ($h=0$), {the solution of Eq.~\eqref{usadelhom}} is a monotonous exponential decay, $f_0\propto \exp\left(-|x|/\xi_N\right)$, while in a homogeneous F an $m=0$ triplet arises through the FFLO effect,\cite{fuldePRL64,larkinJETP65} and the Gor'kov function has two components $f_0$ and $f_z$ describing pair correlations corresponding to the states $\ket{s,m} = \ket{0,0}$ and $\ket{1,0}$.\cite{eschrigPT11} These components exponentially attenuate and oscillate in the F with a characteristic length scale $\xi_F$, $f_{\ket{s,0}}\propto \exp\left(-|x|/\xi_F\right)\,\cos\left(x/\xi_F+s \pi/2 \right)$ ($s=0,1$).

\begin{figure}
\includegraphics[width=0.49\columnwidth]{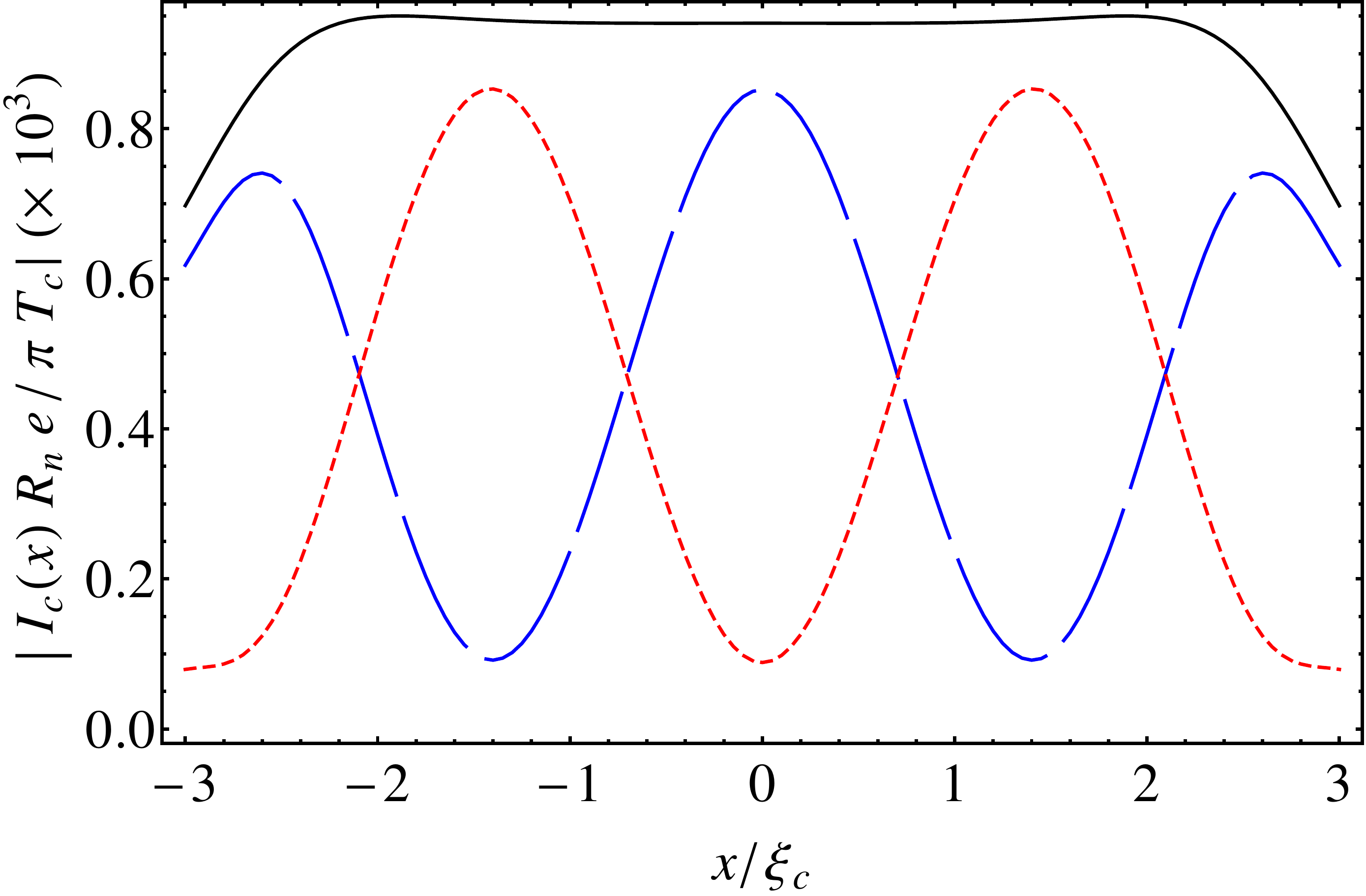}
\includegraphics[width=0.49\columnwidth]{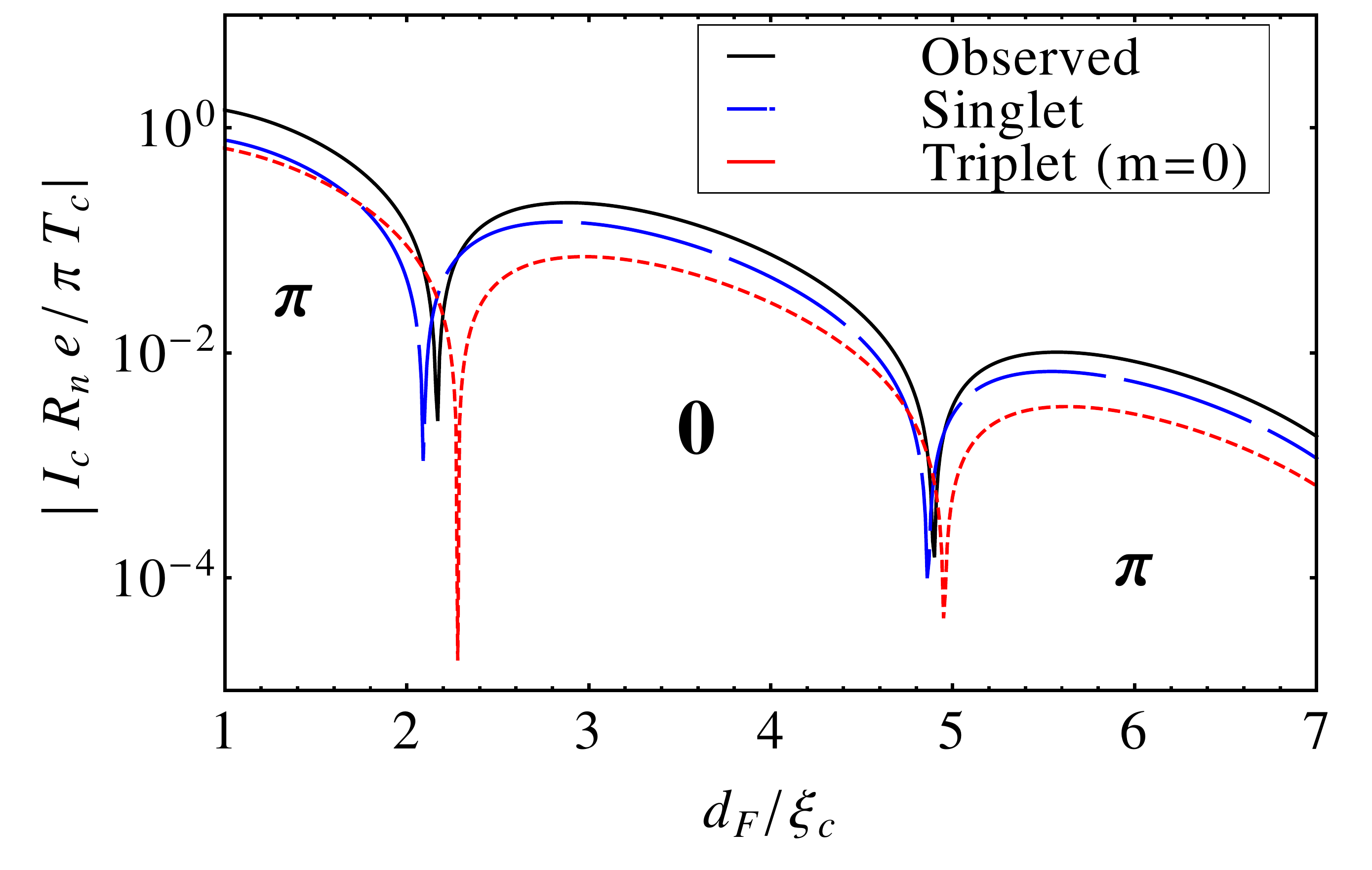}
\caption{\label{fig: jcdens} (color online) Josephson current through an SFS junction with homogeneous F.\cite{bakerEPL14} (a) Josephson current density $I_c(x)$. The decay near the left and right of the system comes from the wide limit approximation.\cite{zaikinZP81} (b) Total current as a function of the F thickness $d_F$.  In both figures we decompose the current in singlet (dashed, blue) and triplet (dotted, red) contributions.  The total (observable) current is plotted as a solid black line and is the sum of these contributions. Parameters used are $h=3\pi T_c$, $T= 0.5T_c$, $\tau\to\infty$, and using the first twenty Matsubara frequencies.}
\end{figure}

The Josephson critical current through a homogeneous F is shown in Fig.~\ref{fig: jcdens}a. We use the decomposition of Eq.~\eqref{IcRNtotdecomp} to demonstrate that even in the homogeneous case, the contributions to the current originating from the singlet and the $m=0$ triplet components ($\alpha = 0,z$) strongly depend on $x$, while the sum of its components $I_c(x)$ is very nearly constant deep in the layer.  The variation of the total current over a length of order $\xi_F$ near the interfaces to the S are due to the wide limit approximation.

To determine the Josephson current, one can either evaluate the current in the middle of the layer (or anywhere, the current density is constant) and multiply by the width of the layer, or integrate the entire current density over the layer. These options give very similar answers. The results of this paper have been obtained through integration over the thickness of the magnetic structure.

Figure \ref{fig: jcdens}b displays the calculated Josephson current for the well studied case of homogeneous magnetic films of different thicknesses using the full, non-linear solution of Eq.~\eqref{usadelhom} from Ref.~\onlinecite{bakerEPL14}. Here again we not only plot the total current (solid line) usually presented but also separate the contributions to the current from the singlet  and triplet correlations.  The interesting feature is that the singlet $I_{c,0}$ and triplet $I_{c,t}$ contributions to the current change phase at slightly different thicknesses, implying that even in homogeneous films the two contributions may compete (for example, the contribution from the singlet may be in a 0 phase while the triplet is in the $\pi$ phase).  We will come back to this scenario to understand the $0-\pi$ transition in the XS (Sec.~\ref{sec: jcurrent}).

\subsection{Usadel's equation for inhomogeneous magnetizations}\label{sec:UsadelIN}

When the magnetization varies in space, superconducting singlet pairs leaking into it will transform into a linear combination of all possible singlet and triplet states $\ket{s,m}$. In particular, whenever the magnetization changes direction long range correlations involving $\ket{1,\pm1}$ states are generated.\cite{bergeretPRL01,kadigrobovEPL01} To include these correlations, it is convenient to consider the Ivanov-Fominov parametrization of the Green and Gor'kov functions\cite{ivanovPRB06,ivanovPRB09}
\begin{eqnarray}\label{eq:parametrization}
\begin{cases}
g_0=M_0\cos\vartheta,\\
f_0=M_0\sin\vartheta,
\end{cases}
\quad\mathrm{and}\quad
\begin{cases}
\mathbf{g} = i \mathbf{M}\sin\vartheta,\\
\mathbf{f} = - i \mathbf{M}\cos\vartheta,
\end{cases}
\end{eqnarray}
Again, the subscripted `0' components relate to the singlet while the vector components are triplet quantities. In the Matsubara representation $M_0$, $\mathbf{M}$ and $\vartheta$ are real functions of position.
Note that the parametrizations in Eqs.~\eqref{usadelhom}
and \eqref{eq:parametrization} involve different ``angular functions'', $\theta= \theta_R + i \theta_I \in \mathbb{C}$ and $\vartheta\in \mathbb{R}$, respectively. 

In the Ivanov-Fominov parametrization, the Usadel equations read ($D=D_F,\, D_S$ is the diffusion constant in the F, S)
\begin{widetext}
\begin{eqnarray}
\label{eq: usadeltheta}
\frac{D}2\nabla^2\vartheta-M_0(\omega_n\sin\vartheta-\Delta\cos\vartheta)-(\mathbf{h}\cdot\mathbf{M})\cos\vartheta&=&0,\\
\label{eq: usadelM}
\frac{D}2(\mathbf{M}\nabla^2M_0-M_0\nabla^2\mathbf{M})+\mathbf{M}(\omega_n\cos\vartheta+\Delta\sin\vartheta)-\mathbf{h}\;M_0\sin\vartheta&=&0,
\end{eqnarray}
\end{widetext}
with the normalization condition
\begin{equation}\label{eq: Mnorm}
M_0^2-|\mathbf{M}|^2=1.
\end{equation}
The functions $M_0$, $\mathbf{M}$ and $\theta$ are odd functions with respect to the Matsubara frequencies $\omega_n$. This can for example be seen by replacing $\omega_n\rightarrow-\omega_n$ in the above equations.

This form of the Usadel equations is not convenient for a numerical treatment.  To achieve that goal we differentiate Eq.~\eqref{eq: Mnorm} twice to obtain 
\begin{eqnarray}
0&=&(\partial_xM_0)^2+M_0\partial_x^2M_0\\
&&-\sum_{\alpha=\{x,y,z\}}\left[(\partial_x M_\alpha)^2+M_j\partial_x^2M_\alpha\right]\nonumber
\end{eqnarray}
 where differential operators are now written explicitly in one dimension assuming that the  magnetization only varies with $x$ across the magnetic structure and is constant in the $yz$ plane. Inserting into Eqs.~(\ref{eq: usadeltheta},\ref{eq: usadelM}) and dotting with $\mathbf{M}$ gives
\begin{widetext}
\begin{eqnarray}
\partial_x^2M_0&=&M_0\left[\sum_{\alpha=\{x,y,z\}}(\partial_xM_\alpha)^2-(\partial_xM_0)^2\right]-\sum_{\alpha=\{x,y,z\}}\frac{2M_\alpha^2}{D_F}(\omega_n\cos\vartheta+\Delta\sin\vartheta)-\sum_{\alpha=\{x,y,z\}}\frac{2M_0M_\alpha}{D_F}h_\alpha\sin\vartheta\label{usadelMrewriteA}\\
\partial_x^2M_i&=&M_i\left\{\left[\sum_{\alpha=\{x,y,z\}}(\partial_xM_\alpha)^2-(\partial_xM_0)^2\right]-\sum_{\alpha=\{x,y,z\}}\frac{2h_\alpha}{D_F}\sin\vartheta\right\}+\frac{2M_iM_0}{D_F}(\omega_n\cos\vartheta+\Delta\sin\vartheta)-\frac{2h_i}{D_F}\sin\vartheta\label{usadelMrewrite}
\end{eqnarray}
\end{widetext}
($i=x,y,z$), where the normalization condition \eqref{eq: Mnorm} has been {used to simplify} some terms. {In this form the} equations are suitable for the numerical relaxation method\cite{numrecipes} thus allowing to solve the full non-linear Eqs.~(\ref{eq: usadeltheta},\ref{eq: usadelM}), without further approximations.  The additional advantage of this parameterization is that the Gor'kov functions can be viewed easily with Eqs.~\eqref{eq:parametrization}.
 
\subsection{Boundary conditions}\label{sec:bc}

The differential equations (\ref{eq: usadeltheta}-\ref{eq: Mnorm}) (or (\ref{usadelMrewriteA}--\ref{usadelMrewrite})) must be supplemented with boundary conditions. There are three types of boundary conditions in the system under consideration. The two first are required for obtaining the Green and Gor'kov functions:\cite{kupriyanov1982influence} 1) Relating the functions in a S and its adjacent F, 2) relating the functions in adjacent Fs. The third set of boundary conditions determines the magnetic state of the system and is generally neglected because the magnetic configuration is considered a given.\cite{bakerAIP16}

Transparent boundary conditions for the Green and Gor'kov functions {between all magnetic layers} are sufficient to capture the physics we consider.  We will match derivatives and values of all the functions at each interface between Fs. On the other hand, the interface between the F and the S has low transparency. \cite{vasenkoPRB08,larkinJETP65} {In this limit, the superconducting pair potential $\Delta$ remains constant up to the boundary of the superconductor in first approximation and} the value of the parameter $\vartheta$ at the SF boundary is set to $\theta_B=\mathrm{arctan}(|\Delta|/\omega_n)$.

In the wide limit of the hybrid system it is possible to treat SF and FS as independent subsystems. At both boundaries we have $(M_0,\mathbf{M})=(1,\mathbf{0})$ and $\vartheta(\mathrm{SF}) = \vartheta(\mathrm{FS}) = \theta_B$. The Green and Gor'kov functions vanish at the other end, that is at the right (left) edge of the SF (FS) part. Numerical evaluations show that in the wide limit setting the latter boundary value of the functions or their derivatives to zero does not change the observable quantities appreciably. As a result of these boundary conditions, we can use Eq.~\eqref{leftright}.

The third set of boundary conditions determines the {\it magnetic} state $\mathbf{h}(x)$ and ensures its stability and these conditions are completely independent from the choices for the correlation functions discussed above. The effect of magnetic boundary conditions are discussed in Ref.~\onlinecite{bakerAIP16}. The main points are that for the helix, $\partial\mathbf{h}(x)/\partial x$ is different at both ends of the magnetic layer except for the special case when the distance is commensurate with the period of the helix.  Thus, in this case the boundary values of the magnetic configuration depend on the thickness of the layer. For the XS $\partial\mathbf{h}(x) /\partial x = 0$ at both edges of the system and this is true for any twist of the magnetization and for any thickness of the XS. In addition, the winding number (the number of times the magnetization rotates by $2\pi$) of the helical structure is unbound and determined by the thickness of the film, while the winding number is less than $1/2$ in all realized XS. The discrete domain wall offers a situation that is intermediate between the two cDWs since we trivially have $\partial \mathbf{h}(x) /\partial x = 0$ as in the XS but the winding number has $N/2$ as upper bound, where $N$ is the number of layers of the magnetic multilayer. The particular aim of a superconducting-spintronic device and the knowledge provided in this paper about the pair correlations in the different systems will determine which of these magnetic boundary conditions is preferred.

\subsection{The magnetic configuration}\label{sec:magconf}

The magnetic profiles discussed in this paper pertain to the two classes of inhomogeneities introduced in Sec.~\ref{s:intro}: \cDW\ or \dDW.  Their distinguishing feature is that in \dDW\ the ``natural" quantization axis (by ``natural" we mean the quantization axis along the magnetic field) changes at a discrete set of well defined positions within the hybrid structure while the \cDW\ is characterized by a continuous rotation of the quantization axis within the magnetic material. The domain wall of the exchange spring can be seen as the continuum limit of a multilayer composed of an infinite number ($N\to \infty$) of misaligned homogeneous Fs of thickness $\Delta x_{F,i} = d_F/N$, with ($i = 1,\cdots, N$). 
As will be seen in the next two sections, cDWs and dDWs generate distinct {mixtures of} pair correlations,  {and in particular} a very different behavior of singlet pair correlations ($\ket{s,m} = \ket{0,0}$).

\begin{figure}
\includegraphics[width=1.05\columnwidth]{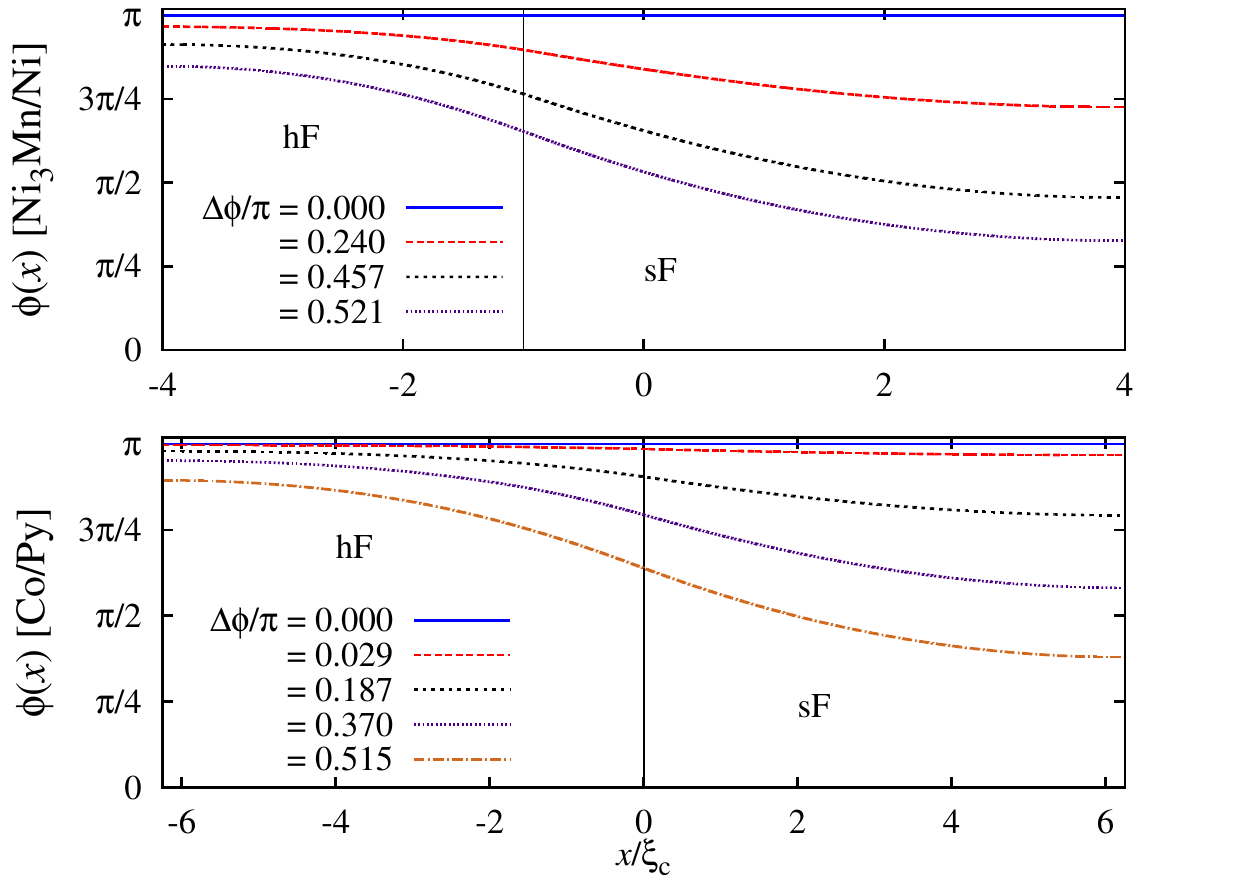}
\caption{\label{XSDWs} (color online) Domain walls in two XSs denoted Ni-XS (top) and Co-XS (bottom) represented by their angle $\phi(x)$, Eq.~\eqref{XSmagnetization}.  The top figure is for the weak magnetization XS Ni$_3$Mn/Ni (hF/sF; see Fig.~\ref{fig:XS}) while the bottom is for the strong Co/Py. Unlike helical structures the XS has flat edges for all twists near the ends of the DW. See text for the parameters.}
\end{figure}
The XS, represented in Fig.~\ref{fig:XS}, allows for a partial to full Bloch domain wall. The magnetic configuration is written as
\begin{equation}\label{XSmagnetization}
\mathbf{h}=- h \sin\phi(x)\mathbf{\hat y} - h \cos\phi(x)\mathbf{\hat z},
\end{equation}
with $|\mathbf{h}(x)| = h$ constant and where $\phi(x)$ is the angle at position $x$ between the magnetization vector and $-\mathbf{\hat{z}}$ (in the present choice of coordinate system). The function $\phi(x)$ has been obtained by minimizing the magnetic energy of the bilayer and provides an excellent description of experimentally realized XS domain walls.\cite{billJMMM04,bakerNJP14}
 In all figures involving a domain wall we characterize the twist of the magnetization by the relative angle $\Delta\phi = \phi({\rm SF}) - \phi({\rm FS})$, which is the angle {between the magnetization vectors}  at the SF and FS interfaces.
In the case of the exchange spring the easy axis is along $\mathbf{\hat{z}}$.

The angle $\phi(x)$ is depicted in Fig.~\ref{XSDWs} for the two XSs considered in this work
and was obtained for the following parameters. The anisotropy energy ratios are $K_h/K_s = 1000$ (Ni-XS), and $625$ (Co-XS). The thicknesses of the layers are $d_{\rm Ni-XS} = t^{\rm Ni}_h+t^{\rm Ni}_s= (2.65 +  5.29)\xi_c$ and $d_{\rm Co-XS}  = t^{\rm Co}_h+t^{\rm Co}_s= (6.25 +  6.25)\xi_c$. The strength of the magnetization are $h_\mathrm{Ni_3Mn}=4.5\pi T_c$, $h_\mathrm{Ni}=4\pi T_c$, $h_\mathrm{Co}=14\pi T_c$, $h_\mathrm{Py}=8\pi T_c$, where $T_c$ is the critical temperature of the Nb/XS proximity system.

The dependence on $x$ is highly non-linear. Furthermore, we note that $\partial \phi/\partial x = 0$ at {the edges of the XS for} arbitrary twists and thicknesses. This implies that the XS domain wall magnetization flattens near the edges of the XS, as seen in the figure;\cite{bakerAIP16} the largest twist remains a  full Bloch domain wall ($\Delta\phi \leq \pi$). This contrasts with the helical magnetic structure, $\phi(x) = Qx$ at fixed $Q$,\cite{bergeretPRB01,wuPRL12,wuPRB12,fritschJPC14,fritschNJP14} where an increase of thickness leads to an unaltered shape of the domain wall (the curvature remains constant) and therefore increased winding, even beyond a $\pi$ rotation of the magnetization. It also contrasts with a magnetic domain wall obtained for a F of infinite thickness that was  overlaid on the magnetic film of finite thickness embedded into a finite size Josephson junction.\cite{linderPRB14} A full discussion is given in Ref.~\onlinecite{bakerAIP16}.

\begin{figure}[h]
\includegraphics[width=\columnwidth]{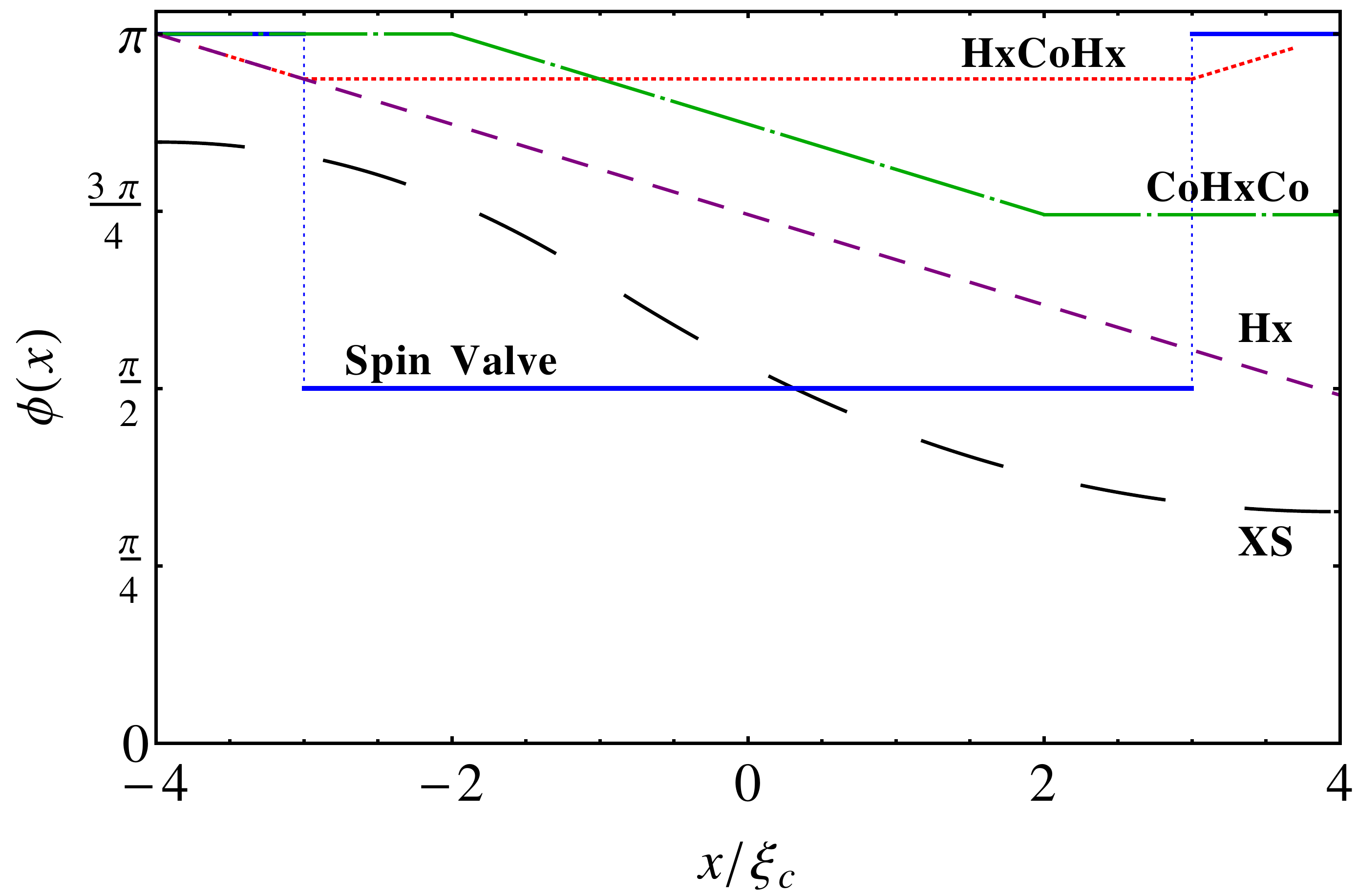}
\caption{\label{DWsdiag} (color online) Comparative representation of the magnetization in \cDW\ and \dDW\  structures studied in this paper. The long dashed black line is for the XS of Fig.~\ref{fig:XS}. The short dashed purple line is for the helix (Fig.~\ref{helix}), and the solid blue lines are for the spin valve structure of Fig.~\ref{spinvalvediagram}. We also show the magnetization profile of Hx/Co/Hx and Co/Hx/Co structures where Hx is a helix described by Eq.~\eqref{hhelical} and are similar to Ref.~\onlinecite{robinsonS10} except that $Q\xi_c\lesssim 1$ is here much smaller than the case of Ho ($Q\xi_c = 11$). Parameters are those of the Ni-XS for the XS, $Q\xi_c = 0.2$ for the Hx, and $\phi = \pi/2$ for the spin valve.}
\end{figure}

The different magnetic configurations considered in this paper are depicted in Fig.~\ref{DWsdiag}. One recognizes the non-linear feature of the XS domain wall and how it is unique. {Its features have similarities with each of} the other structures, but none models the XS. In particular, the edges of the XS resemble the misaligned homogeneous spin-valve but clearly not the helical structure. Conversely, the middle part of the XS resembles the helical structure but is different from the middle layer of the spin valve.

\section{Pair Correlations in an exchange spring}\label{sec: SXSS}

Using the formalism of the previous section we now undertake a comparative study of pair correlations in different hybrid systems. 
The magnetic multilayers considered here are very wide (several tens of $\xi_F$). According to common understanding the presence of ``short range" $\ket{s,m} = \ket{s,0}$ components should thus only be visible near the SF and FS interfaces. As shown here this is not the case for \cDW\ systems such as the XS.

In the next subsections, we first consider triplet correlations $\mathbf{f}$ then analyze the relation between $m=0$ and $m=\pm 1$ components usually termed short and long range components. We also first discuss \cDW\ (XS, helixes, etc.) systems then compare to the corresponding functions in \dDW\ (spin valves\cite{houzetPRB07}) systems.
The figures only display the pair correlations in the magnetic material and assume that the S is located on the left or the right of the figure; the left {(right)} boundary of the figures thus coincide with the SF {(FS)} interface. 

\subsection{Triplet Components}\label{sec: LRTC}
\begin{figure*}
\includegraphics[width=2\columnwidth]{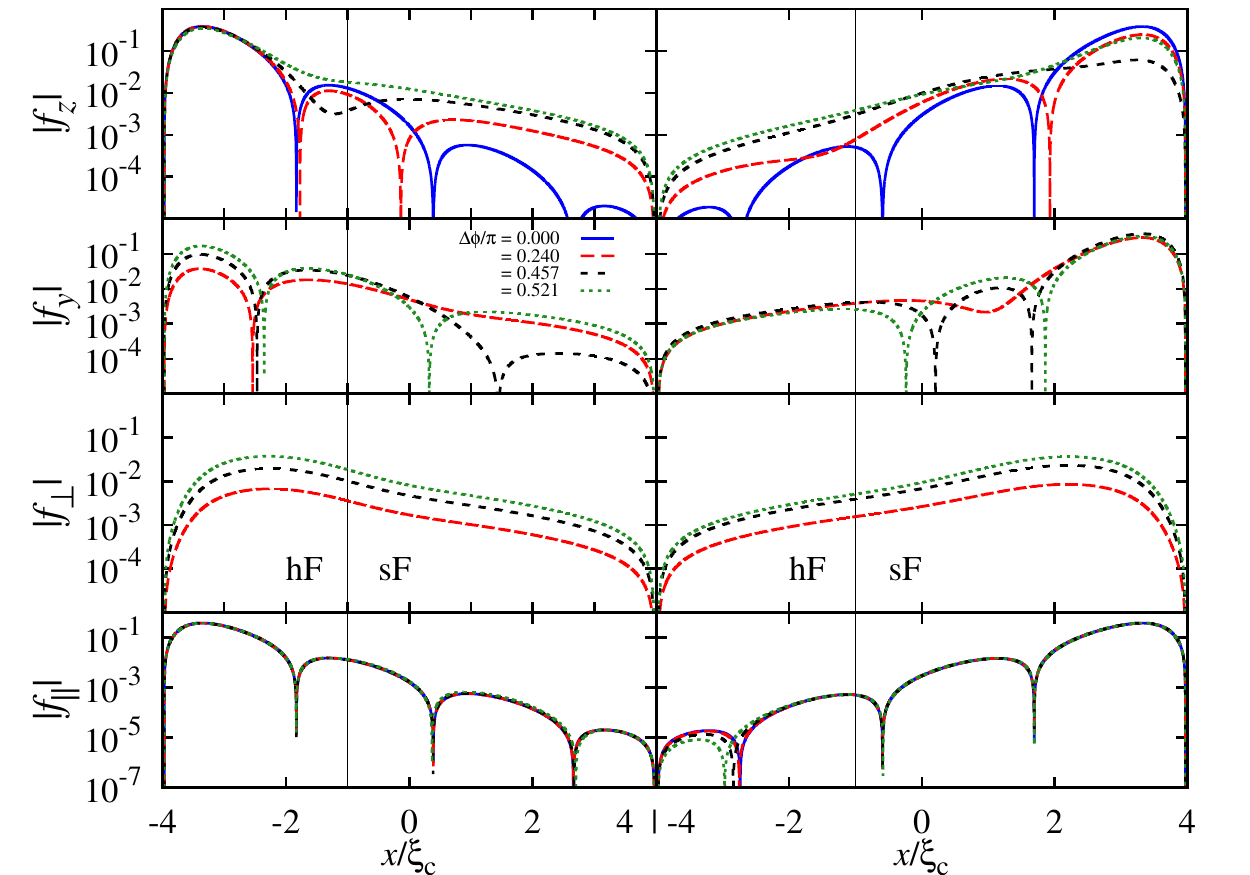}
\caption{\label{varNiGorkov} (color online) Components of the Gor'kov triplet vector function $\mathbf{f}$ in Ni$_3$Mn/Ni, represented in the Cartesian basis $\{\mathbf{e}_y,\mathbf{e}_z\}$ (first two rows) and the rotating basis $\{\mathbf{e}_\parallel,\mathbf{e}_\perp\}$ (two last rows; see Eq.~\eqref{rotatingbasis} and text). The left (right) column depicts the Gor'kov functions for S located on the left (right) of the XS. In each figure the curves are obtained by solving Eqs.~(\ref{eq: usadeltheta}, \ref{eq: usadelM}) for the twists of Fig.~\ref{XSDWs}. Oscillatory sections (dips) of the functions denote regions where $m=0$ triplet components dominate while longer decays are $m\neq0$ components.  In the first two rows, one observes the tuning in of the long ranged components with increasing twist, recognizable with the characteristic long exponential tail and concomitant disappearance of the oscillations. The last two rows represent the same Gor'kov vector function $\mathbf{f}$ in the rotating basis displaying the clear separation of short, oscillating ($f_\parallel$) and long ($f_\perp$) range components. Parameters are: $K_h/K_s=1000$, $t_h \approx 2.65\xi_c$, $t_s \approx 5.29\xi_c$, $h_h=4.5\pi T_c$, $h_s=4\pi T_c$, $T= 0.2T_c$.}
\end{figure*}

\begin{figure*}
\includegraphics[width=2\columnwidth]{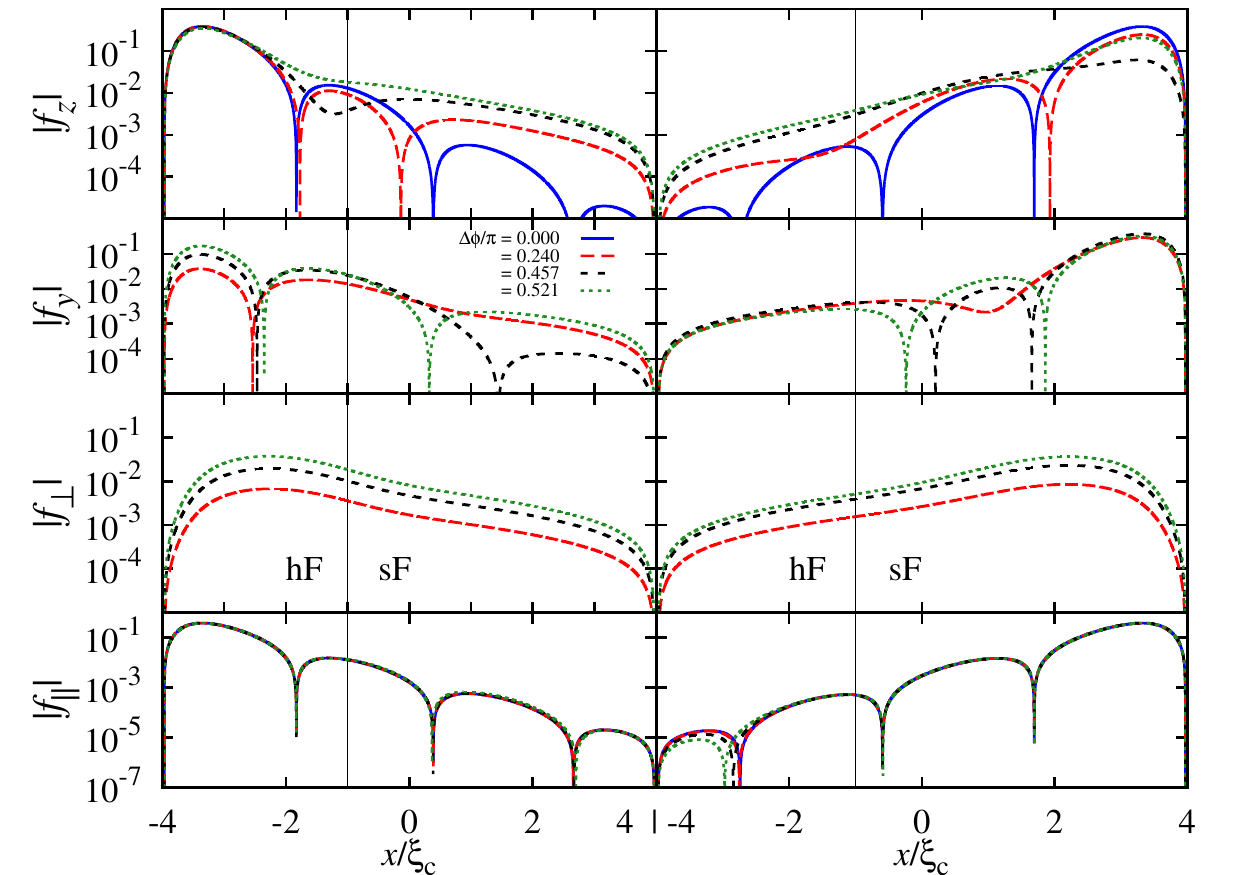}
\caption{\label{varCoGorkov}(color online) Similar to Fig.~\ref{varNiGorkov} but for Co/Py. Upper two rows: Gor'kov functions $f_y$, $f_z$ for the S/XS (left column) and XS/S (right column) with different twists of the XS. Lower two rows: The same Gor'kov function $\mathbf{f}$ in the rotating basis, Eq.~\eqref{rotatingbasis} (see text). The text discusses the marked change from short to long range behavior of $f_\parallel$ far away from the S/XS and XS/S interfaces. Parameters are: $K_h/K_s=625$, $t_h = t_s\approx6.25\xi_c$,  $h_h =14\pi T_c$, $h_s=8\pi T_c$, $T= 0.2T_c$.}
\end{figure*}

The components of the Gor'kov vector $\mathbf{f}(x)$ (triplets) for different twists $\Delta\phi$ of the exchange spring magnetization are shown for the Ni$_3$Mn/Ni exchange spring in Fig.~\ref{varNiGorkov} and for the Co/Py exchange spring in Fig.~\ref{varCoGorkov}. The scalar (singlet) component $f_0$ of these same systems is discussed in Sec.~\ref{sec: LRSC}. The Ni-based XS (henceforth referred to as the Ni-XS) is made of materials with weak magnetization $|\mathbf{h}(x)| = h$ and large magnetic energy anisotropy ratio while the Co-based XS (or Co-XS) has strong magnetization and weaker anisotropy ratio.  The important distinction between these systems for the purpose of this work is that the Ni-XS is a weaker pair-breaker, allowing pairs to diffuse farther into the magnetic structure.

Consider the Gor'kov functions of Fig.~\ref{varNiGorkov}. For a homogeneous magnetization ($\Delta \phi = 0$) the $m=0$ triplet component (solid blue line) only appears along the $\mathbf{\hat z}$ direction; $f_y = 0$.  This is expected for a singlet Cooper pair superconductor placed on either side.  The component displays the well-known oscillations  that are ultimately responsible for the Buzdin-Bulaevskii-Panyukov $0-\pi$ transition of the Josephson current (with a full discussion delayed until Sec.~\ref{sec: jcurrent}). The oscillatory fast decay with characteristic length $\xi_F$ is identified in the literature as the ``short-range'' component of the pair correlations.  Inducing a small twist, $\Delta\phi\gtrsim 0$ (for example the dashed red line), the {$m\neq 0$} components appear (here as an $|f_y|$ component) throughout the layer with a very long and slow non-oscillatory decay{, coined ``long-range'' correlations in the literature}.  {Increasing} the twist ($\Delta\phi = 0.24, 0.46, 0.52$ in Fig.~\ref{varNiGorkov}) the two nodes in $|f_z(x)|$ that are closest to the SF interface move towards each other and disappear, implying the vanishing of the oscillatory behavior and leaving only a long range component. At the same time, the long range behavior of $|f_y(x)|$ gradually disappears and nodes (oscillations) appear. Thus, as one component ($f_z$) transforms from a short- into a long-range component the other component ($f_y$) does the opposite as $\Delta\phi$ is increased.

The behavior just described reflects the continuous rotation of the quantization axis from $\hat{\mathbf{z}}$ towards $\hat{\mathbf{y}}$, as  demonstrated in the bottom two rows of Fig.~\ref{varNiGorkov} where we plot the components of the Gor'kov vector, $|f_\perp|$, $|f_\parallel|$, in the basis rotating with the magnetization. The component perpendicular to the magnetization has a non-oscillatory weak decay at all twists, contrasting with the parallel component. We  {also note} that for low twists of the magnetization the $m=0$ triplet component, $|f_\parallel|$, dominates while for stronger twists the $m\neq 0$ components, $|f_\perp|$, dominate. This explains the growing Josephson current observed with increasing $\Delta\phi$ in Ref.~\onlinecite{bakerNJP14} -- the ``long ranged" pair correlations are increasing in amplitude.\cite{bergeretPRL01,kadigrobovEPL01}  Some change in the $f_\parallel$ components appear far from the SF interface (right column of Fig.~\ref{varNiGorkov}) that is due to  the cascade effect \cite{bakerEPL14}  This will be revisited {below} when discussing the Co-XS.

Observe in Fig.~\ref{XSDWs} that, with growing $\Delta\phi$, the curvature of the magnetization twist increases near the edges. This leads to a decrease of the width near the SF and FS interfaces where the magnetization is approximately constant, approaching values closer to $\xi_F$ and thus allowing for the formation of a larger $m\neq0$ component to develop. This is evidenced by the increase of $|f_y|$ or $|f_\perp|$ with increased $\Delta\phi$. The increased twist is expected to lead to a stronger current.\cite{bakerAIP16}

We note two more features seen in Fig.~\ref{varNiGorkov}. The Gor'kov functions are smooth across the magnetic bilayer, reflecting the transparent interface conditions (located at $x=-1.32\xi_c$) chosen for this study. Finally, we point out that the Gor'kov functions calculated for the SF and the FS interfaces (left versus right panels in Fig.~\ref{varNiGorkov}) are not symmetric since the hard and soft layers constituting the XS have different magnetic properties; the XS is an asymmetric bilayer with a magnetic interaction across the interface.

\subsection{{Mixing of $m\neq 0$ and $m=0$ triplet components}}\label{sec: LvS}

A major difference between the continuously rotating magnetization of a domain wall (cDW) and misaligned homogeneous Fs (dDW) is the behavior of $m=0$ (usually termed ``short range") components deep in the magnetic material.\cite{bakerEPL14} Each rotation of the magnetization can be interpreted as a rotation of the quantization axis, resulting in a new linear combination of the states $\ket{s,m}$. This is related to the effective boundary conditions introduced in Ref.~\onlinecite{bakerEPL14}. In the case of a \dDW\ (see Sec.~\ref{sec:hybridsystems} for more details) this remix {may only occur} at the discrete set of interfaces, between various {magnetic} layers. Away from these interfaces the $m=0$ contributions decay exponentially over the short-range scale $\xi_F$. Contrastingly, in a \cDW\ system the remix of components occurs at {\it all} points across the continuously rotating magnetization. The redistribution of weight between the various components $\ket{s,m}$ (the cascade effect of Ref.~\onlinecite{bakerEPL14}) effectively reduces the decay of the singlet and $m=0$ triplet components in the magnetic material. While the abrupt rotation of the magnetization in a dDW causes a sizable redistribution of pair correlations as seen in Ref.~\onlinecite{bakerEPL14}, the cDW provides for gentler effects.

Fig.~\ref{varCoGorkov} explicitly shows the mixing of components with the continuous rotation of the Co-XS.
The figure displays the same Gor'kov functions as in Fig.~\ref{varNiGorkov} but for the Co-XS with stronger pair breaking and weaker anisotropy ratio. We observe similar, though more dramatic changes of the Gor'kov function $\mathbf{f}$ than in the Ni-XS.

The most dramatic feature of the \cDW\ is best seen in the rotating basis (bottom rows of Fig.~\ref{varCoGorkov}). We note that {deep in the magnetic material} the $m=0$ triplet component (seen in $|f_\parallel |$) does not decay in an oscillatory exponential way to zero but saturates. This component undergoes a marked change of character far away from the SF (or FS) interface ({\it e.g.} $x\in[2,6]\xi_c$ in the left column {bottom row of Fig.~\ref{varCoGorkov}}): $f_\parallel$ morphs into a {\it long} range component and is {\it in addition} to the long range component observed in $f_\perp$. Even though the value of this $m=0$ component is 10$^3$ times smaller than the $m\neq 0$ component the saturation effect is unexpected and is several orders of magnitude larger than the usually anticipated behavior. 

The slowly decaying feature (at $x>2\xi_c$) in $f_\parallel$ is understood with the cascade effect in a continuous rotation of the magnetization.\cite{bakerEPL14} Noteworthy is that this effect is present throughout the cDW but is only visible far enough from the interface.
{The reason is that} there are two contributions to the Gor'kov function. One is due to the {superconducting Cooper pairs leaking into the magnetic material. The $m=0$ singlet and  triplet components generated at the SF interface simply oscillate and decay exponentially with lengthscale $\xi_F$.}  The other part comes from the rotation of the magnetization and the related cascade effect generating the $m=\pm 1$ components. The magnitude of that part and its decay are much weaker and the signal does not oscillate. {Thus, only when the first component has died off, does the second component become dominant.} This heuristic interpretation is motivated by the comparison of the Co-XS and Ni-XS {and from the study of a helical structure in Ref.~\onlinecite{bergeretPRB01}.}  Although the contribution to $f_\parallel$ is also present in Ni-XS, it is not seen in Fig.~\ref{varNiGorkov} because the XS has weak magnetization, implying that $\xi_F$ is larger and the cascade effect remains buried hence giving only a small correction to the known oscillatory behavior.
That the contribution from the superconducting singlet pairs leaking into the magnetic material masks the cascading contribution close to the SF interface means that the cascade effect is relatively weak in the \cDW.

The behavior just described is also visible in the cartesian basis where both $f_y$ and $f_z$ show non-oscillatory long-range features. One common feature between the Co-XS and the Ni-XS systems is found in $f_y$. This component contributed from the left S undergoes a change {of sign}  deep in the magnetic layer at $x\approx2\xi_c$. This implies that {the oscillation characteristic of} `short' ranged behavior can appear far from the SF interface.\cite{bakerEPL14}

The outcome of this analysis is that although the decay length of the $m=0$ components is associated with $\xi_F$ and therefore short range, the cascade effect in a continuous rotation of the magnetization fuel these components to mimic an $m\neq0$ Gor'kov function with a long decay length. The distinction between long and short ranged components is therefore not as clearly established in a continuously rotating magnetization (\cDW) as in a multilayer of misaligned homogeneous Fs (\dDW). 

\subsection{Persistent singlet components}\label{sec: LRSC}

\begin{figure}
\includegraphics[width=\columnwidth]{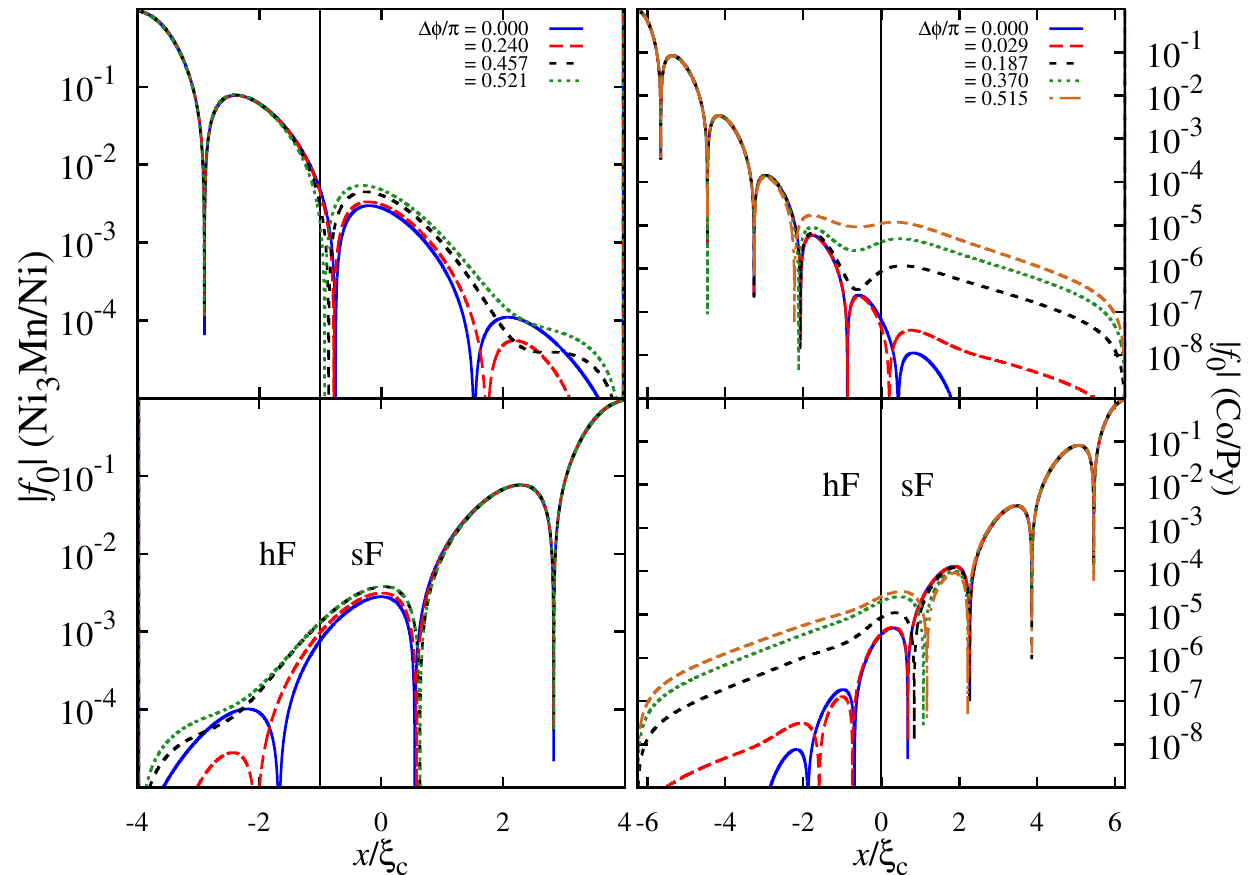}
\caption{\label{Gorkovf0XS}(color online)  Singlet Gor'kov function $|f_0|$ for various twists in a weak XS, Ni$_3$Mn/Ni (left column) and a strong XS, Co/Py (right column). The first (second) row is for an SF (FS) system. Noteworthy is the increase of the singlet component deep in the XS. This behavior is reminiscent of a long-range behavior but results from the cascade effect in a continuously rotating magnetization. The singlet component acquires the same characteristic long range appearance as the $m\neq0$ component (see text). Parameters are identical to those of Fig.~\ref{varNiGorkov} and \ref{varCoGorkov}, respectively. Note the different range of the ordinates scale.}
\end{figure}

Having established the existence of notable contributions of the $m=0$ triplet components resulting from the cascade effect in a \cDW\ we analyze features of the singlet part. Figure~\ref{Gorkovf0XS} shows the Gor'kov function $|f_0(x)|$ for the Ni-XS (left column) and the Co-XS (right column).  The singlet component clearly increases by several orders of magnitude and loses its oscillatory feature deep in the magnetic material when increasing the twist of the domain wall, just as the $f_\parallel$, {$m=0$ triplet} component did for the Co-XS.  {This behavior of the singlet pair correlations is very unexpected. Singlet correlations are determined by the scalar function $f_0$ that should only be sensitive to the magnitude of the magnetization and should not depend on its direction! Only the triplet correlations described by the vector $\mathbf{f}$ are expected to respond to the rotation of an external field.}  
Because the XS is very wide and the decay of $|f_0|$ is reduced with increasing magnetic twist only far from the SF or FS edge, it cannot be ascribed to the increase one would expect from mimicking antiferromagnetism through twisting the domain wall. The XS is so wide that this increase cannot come from the mimic of the antiferromagnetic state.\cite{bergeretPRL01b}  Rather, the cascade effect and its related reverse FFLO effect are at work converting the $m\neq0$ components to $m=0$ components and generating, at the same time, a singlet component.\cite{bakerEPL14}

The effect is exacerbated in the stronger ferromagnet Co-XS though the absolute value of the component is smaller.  This effect is indicative of a significant change in the physics of the pair correlations as compared to that usually presented and that is only observed in \cDW\ materials.  The rotating magnetization provides a channel to generate singlet components and one therefore finds singlets well beyond one coherence length. We emphasize that this singlet component always involves the {concomittance}  of $m=0$ and $m\neq 0$ triplet components. 

 {Following similar reasoning as in the previous section,} the results of Fig.~\ref{Gorkovf0XS} invite us to distinguish {singlet pair correlations due to the proximity effect (singlet Cooper pairs leaking into the magnetic material) and singlet components due to the cascade effect (generation of $m=0$ component in the rotating magnetization).}  As in Fig.~\ref{varCoGorkov} the two singlet components can be identified in Fig.~\ref{Gorkovf0XS} through the change of functional behavior of the Gor'kov function. {The second contribution dominates far from the interface, when the first  decayed sufficiently.} 

Finally, we note that the long ranged singlets suggested to exist in nanowires are different from the effect discussed here.\cite{konschellePRB10} Our singlet component remains short ranged -- the characteristic length of the decay is $\xi_F$ -- but reemerges as a consequence of the cascading effect and the reverse FFLO effect  {that} sustains the production of $m=0$ components {in a cDW}. 

The results presented here for the singlet Gor'kov function confirm the statement made in the previous section, that the distinction between so-called `short' and `long' ranged components is somewhat ill-defined in \cDW\ materials.  In continuously rotating magnetization it is more instructive to categorize the components by their quantum numbers $\ket{s,m}$,  as we do in this work.  The exchange spring represents an instance of \cDW\ systems where a cascade of components is present in the magnetic material and has measurable consequences as discussed in Ref.~\onlinecite{bakerNJP14} and a later section.

\section{Pair correlations in other hybrid systems}\label{sec:hybridsystems}

In the previous section, we discussed pair correlations in  {XSs} by displaying all components of the Gor'kov functions, $f_\alpha$, with $\alpha = 0,y,z, \perp, \parallel$.  We compare here these results with the pair correlations found in other hybrid structures discussed in the literature. In particular, we investigate whether the features seen in the XS are also seen in other \cDW s and how they contrast with structures belonging to the \dDW\ category. To this aim, we show the Gor'kov functions for continuous helical domain walls pertaining to the \cDW\ category, and multilayers of homogeneous, misaligned Fs of the \dDW\ category.

\subsection{Helical Domain Walls}\label{ss:helicalDW}

First, we examine the helical domain wall as it is closest to our XS domain wall.  For the helical magnetization we replace the non-linear function of position $\phi(x)$ of the XS in Eq.~\eqref{XSmagnetization} by a linear dependence on position $\phi(x)  = \pi - Qx$ leading to a magnetization of the form\cite{bergeretPRB01,wuPRL12,fritschNJP14}
\begin{equation}\label{hhelical}
\mathbf{h}_\mathrm{helical}= - h\sin (Qx) \mathbf{\hat y} + h\cos (Qx) \mathbf{\hat z}.
\end{equation}

\begin{figure}
\includegraphics[width=\columnwidth]{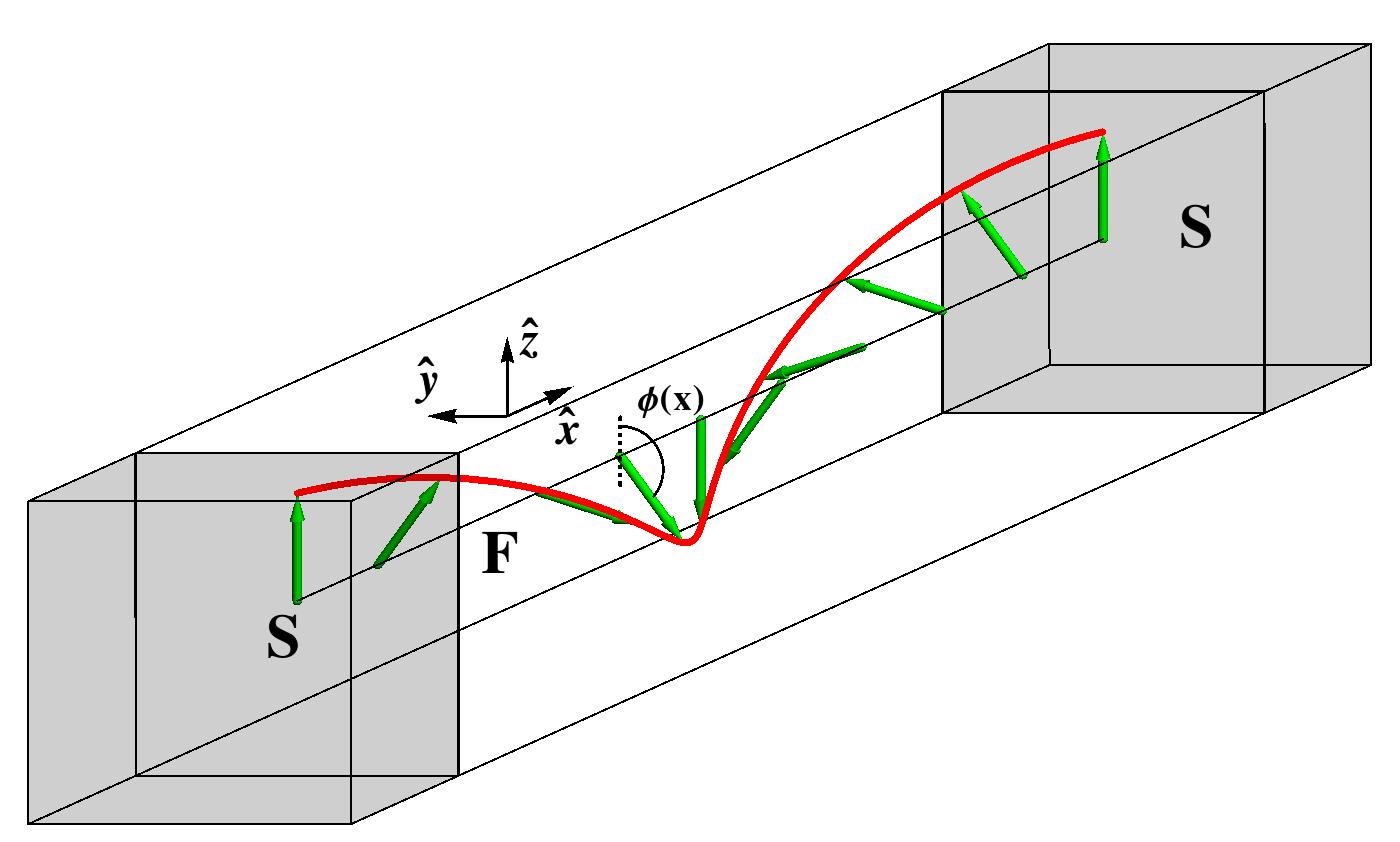}
\caption{\label{helix} (color online) Pairing types in a helical domain wall. This represents for example the $yz$ projection of the Holmium magnetization (canted $10^\circ$ towards $\hat{\mathbf{x}}$). Shown is the layer thickness for which the winding of the domain wall is $2\pi$, but often thicknesses are studied corresponding only to partial domain walls.} 
\end{figure}
{Figure~\ref{helix} depicts the case of a full rotation. This expression relates to the conical magnetization studied in Refs.~\onlinecite{bergeretPRB01,wuPRL12,wuPRB12,fritschNJP14,fritschJPC14}.\cite{helixanglenote} The results in these papers indicate that  for the physics discussed here there is little difference between the cases $\alpha = 90^o$ (with respect to the $x-$axis) chosen here and $\alpha = 80^o$ found in the conical magnetic profile of Holmium.

Figure~\ref{fhelical} displays the Gor'kov functions for the helical structure of Eq.~\eqref{hhelical} with three examples that are close to the configurations discussed in Refs.~\onlinecite{bergeretPRB01,fritschNJP14}. Note, however, that Ref.~\onlinecite{fritschNJP14} discusses the opposite, clean limit case. All panels show three different curves for three different twists at fixed thickness of the F. The $Q\xi_c = 0.001, 0.01$ twists are weak while $Q\xi_c = \pi/16$ corresponds to a full Bloch domain wall over the F ($\Delta\phi = \pi$).

We point out similar trends but also notable differences between the helical structure and the XS. We emphasize first that the three curves shown in each panel of Fig.~\ref{fhelical} {\it cannot} be found in the same material since a given system has a fixed value of $Q$ (for example, Holmium has $Q\xi_c\approx 11$). Thus, in stark contrast to the XS figures \ref{varNiGorkov}-\ref{Gorkovf0XS} where all curves are obtained with the same system, here we are comparing the pair correlations for helical structures of different materials.

The left (right) column shows the pair correlations when the superconducting electrons leak into the helical F from the left (right) of the figure. The lack of mirror symmetry of the $f_y$ and $f_z$ curves with respect to a vertical plane parallel to the layers has a different origin than in the XS.  The XS being made of two Fs with different magnetic properties the electrons enter a different material when penetrating the XS structure from the left or right which causes different decays of the correlations.  By contrast a helical F is composed of one material but the magnetic boundary conditions (the curvature in particular) change with $Q$.\cite{bakerAIP16} {The boundary condition also changes at given twist when varying the thickness of the material.}  While $\phi(x)$ flattens at both edges of the XS (leading to an infinite curvature on both ends), the corresponding function in the helical structure has different slopes at the right edge as one varies either $Q\xi_c$ or the thickness.\cite{bakerAIP16}
{Noteworthy is that the} parallel and perpendicular components, $f_\parallel$, $f_\perp$, are symmetric, reflecting the linear form of the {angle and the constant curvature of the} helical profile.  The XS does not  possess this feature.

The {uppermost} row shows the singlet pair correlations $|f_0|$. We note, as in the XS, the presence of long range singlet {correlations}  emerging with increasing twist $Q\xi_c$ resulting from the cascade effect. The following two rows display features of $\mathbf{f}$ that are similar to the XS: The presence of a twist (inhomogeneity) generates $m\neq 0$ triplet components that increase in magnitude as one increases the twist. 
The intermixture of all components is most evident for the full domain wall ($Q\xi_c=\pi/16$) since the curves displays features from both; the long decay tail is indicative of $m\neq0$ components and the oscillations reflect the presence of $m=0$ terms.

\begin{figure}
\begin{center}
\includegraphics[width=\columnwidth]{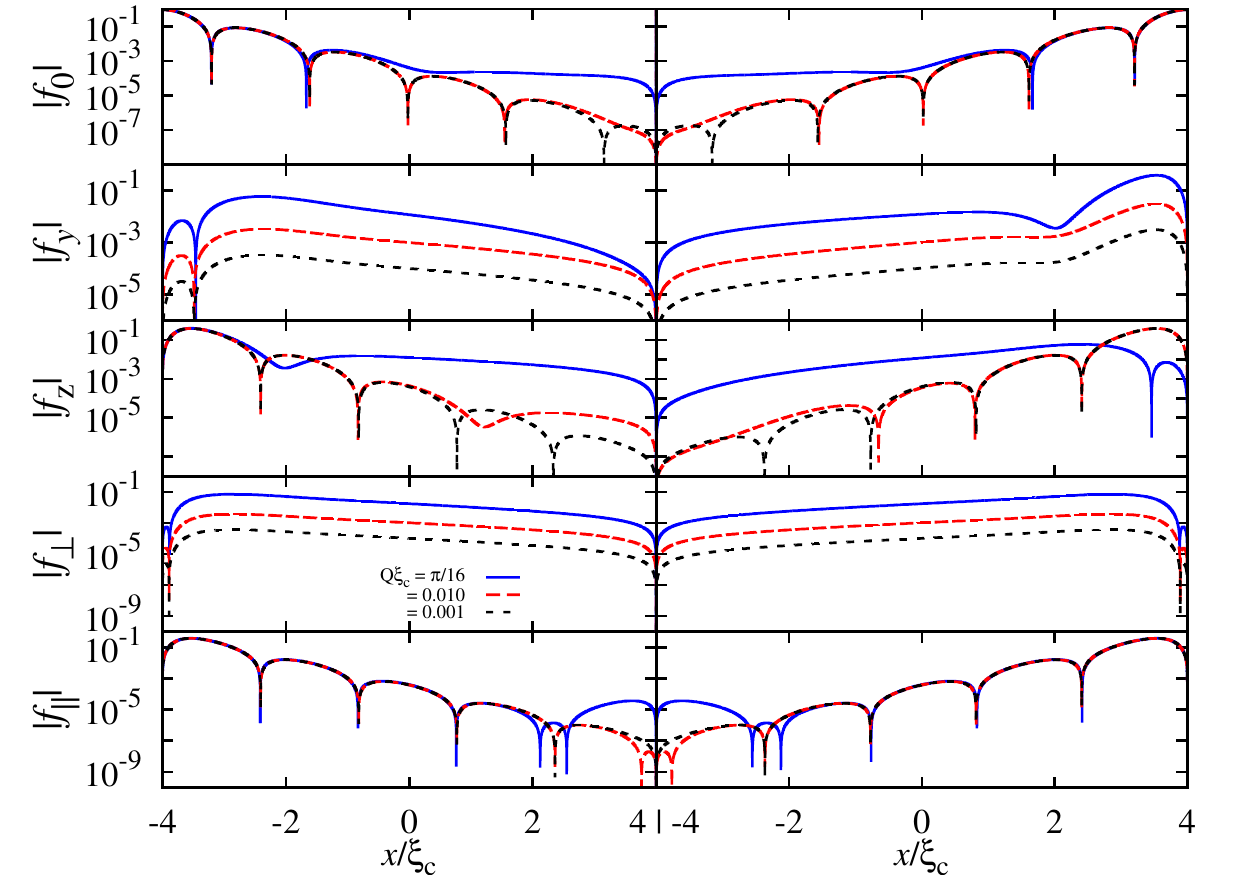}
\end{center}
\caption{\label{fhelical}(color online) Singlet and triplet Gor'kov functions for the helical magnetic structure, Eq.~\eqref{hhelical}, with $Q\xi_c = 0.001, 0.01, \pi/16$, for the SF (left column) and FS (right column) system. The $\pi/16$ case corresponds to $\Delta\phi/\pi=1$ (full Bloch domain wall). The figure and its relation to the XS are discussed in the text. We emphasize that in contrast to the XS of the previous figures the different values of $Q$ imply results for {\it different} materials. Parameters used are $T= 0.2T_c$, $h=8\pi T_c$, and $d_F=8\xi_c$.}
\end{figure}

The {two} lower rows of Fig.~\ref{fhelical} present the Gor'kov functions in the rotating basis, Eq.~\eqref{rotatingbasis}. One observes the same general behavior as for the XS case with substantial twist of the domain wall. {The}  increase {seen in $f_\parallel$ near the edge of the F opposite to the interface with the S}  stems from the cascade effect; some {$m=0$ components are regenerated by the continuously rotating magnetization of the cDW.} 
 The main statement {made in the XS} is confirmed: the {cDW} generates {$m\neq0$ {\it and} $m=0$} components throughout the cDW.\cite{bergeretPRB01} 

We note in fact that the $m\neq 0$ components are stronger {in the helical structure} when compared to the XS.  This is also related to the different magnetic profiles at the edges of the systems. 

Another difference between the XS and the helical structure is observed very close to the SF or FS edge (see for example the fourth row representing $f_\perp$). We note a node and thus a change of sign of the Gor'kov function very near the edge. In contrast, the pair correlation of the XS has always the same sign in this vicinity.  This results from the nonlinearity of the XS magnetic profile {angle}  $\phi(x)$.\cite{bakerAIP16}   

Finally, we point out that Holmium used in experimental setups \cite{robinsonS10} has a very strong helical twist, corresponding to large values of $Q$ ($Q\xi_c \approx 11$ in our units). Thus, Ho/Co/Ho or Co/Ho/Co layers used in experiment are actually more related to the class of \dDW\ or to a spin active interface\cite{eschrigPT11} than a continuous \cDW.

\subsection{Discrete Domain Walls}\label{ss:discreteDW}

The other inhomogeneity studied extensively in the literature is that of misaligned homogeneous ferromagnetic layers where the magnetization changes direction at a discrete set of points in the multilayer (at the interfaces) and thus belongs to the \dDW\ class of systems. Various combinations have been studied and we consider here the case closest to the XS and schematically depicted in Fig.~\ref{spinvalvediagram}, namely three misaligned homogeneous layers F$_1$F$_2$F$_3$, as studied in Ref.~\onlinecite{houzetPRB07}.
\begin{figure}[h]
\includegraphics[width=\columnwidth]{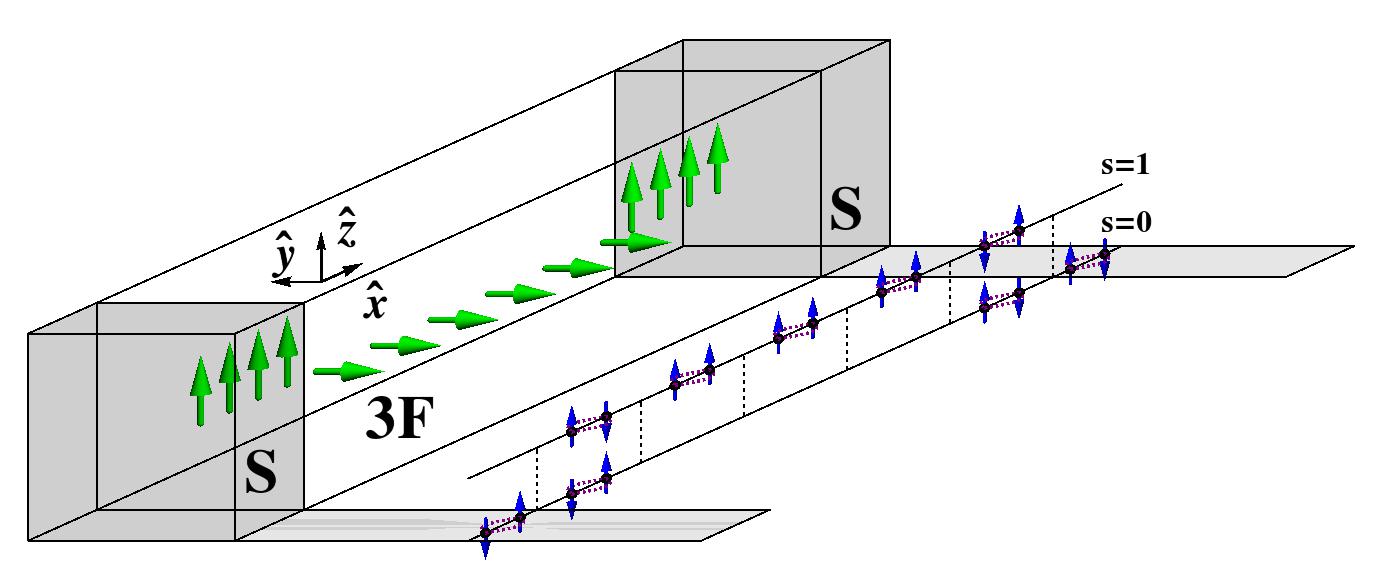}
\caption{\label{spinvalvediagram}(color online) Schematics of the SF$_1$F$_2$F$_3$S spin valve structure.  Thick arrows (green) denote the direction of magnetization in the three F layers while paired spins (small arrows) denote the predominant pairing types.}
\end{figure}
\begin{figure}[h]
\includegraphics[width=\columnwidth]{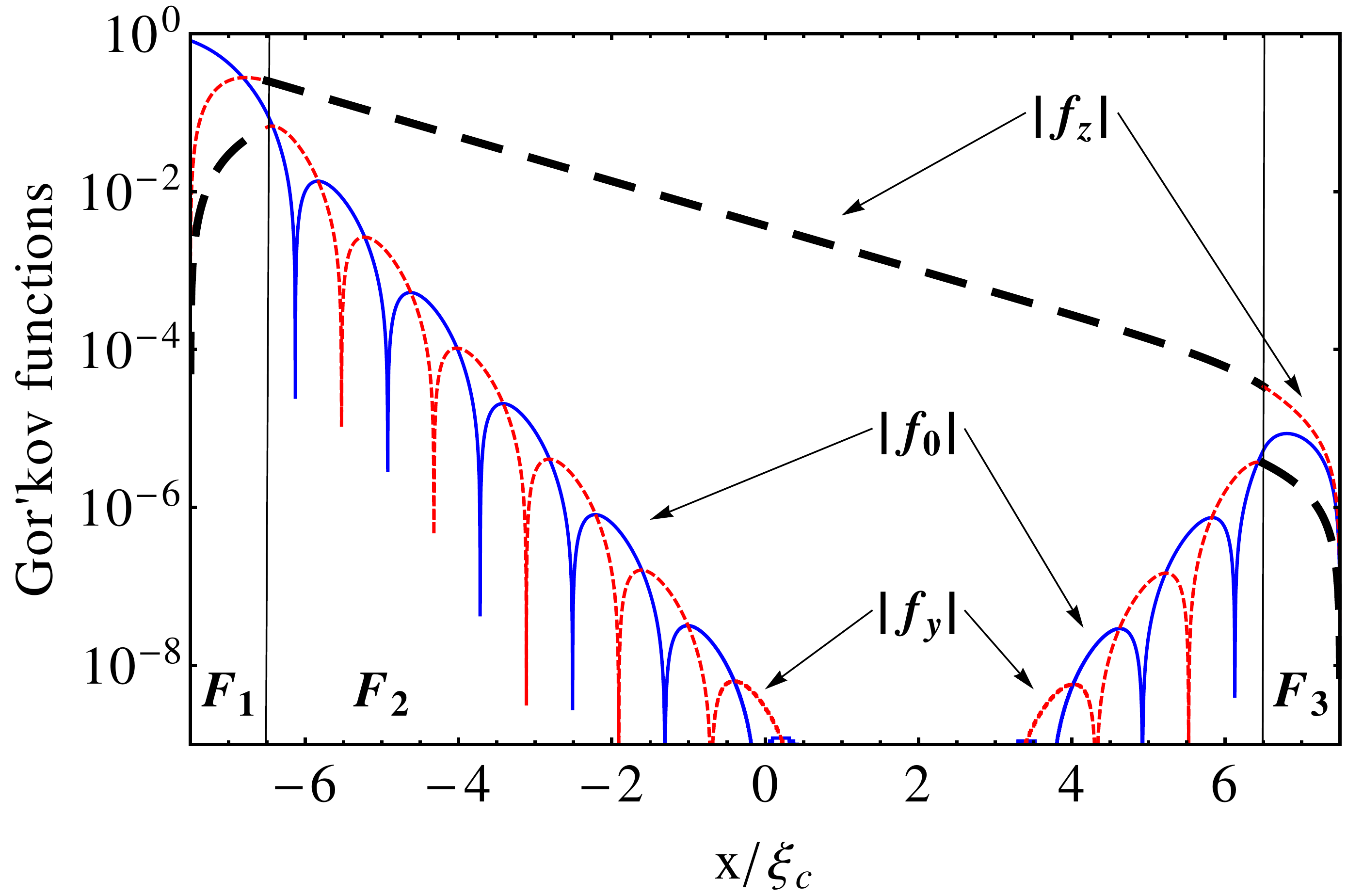}
\caption{\label{S3FSgorkov} (color online) Gor'kov functions for the spin valve (SF$_1$F$_2$F$_3$S) of Fig.~\ref{spinvalvediagram} {obtained for singlet Cooper pairs leaking from the left S only}. Solid lines are singlets (blue), dotted lines are $m=0$ triplets (red), and thick dashed lines are $m=\pm1$ triplets (black).  Parameters used are $h_{F_1,F_3}=3\pi T_c$, $h_{F_2}=14\pi T_c$, $T= 0.4T_c$, $d_{F_1F_2F_3}=15\xi_c$, and $d_{F_1,F_3}=\xi_c$.}
\end{figure}
Figure~\ref{S3FSgorkov} explicitly shows the Gor'kov functions for three layers of same saturated magnetization and thicknesses satisfying the relations $d_{F_1},d_{F_3} \sim \xi_F \ll d_{F_2} \ll \xi_S$. We thus show a system with very wide middle layer, in the spirit of Ref.~\onlinecite{houzetPRB07}, to clearly view the behavior of the Gor'kov functions. 

We emphasize that all Gor'kov components are continuously differentiable functions in the entire multilayer. Contrary to all other figures in this paper the dashed and dotted line types (red and black color code) used in Fig.~\ref{S3FSgorkov} denote the $m=0$  and $m\neq 0$ pair correlations, respectively. For example, $f_z$ is depicted as a dotted (red) line in $F_1$ and $F_3$ where it describes $m=0$ correlations ($f_z$ is the component parallel to the magnetization {in these layers}), while {this same component} $f_z$ is represented with a dashed (black) line in $F_2$ to indicate that it represents $m\neq 0$ correlations in that layer ({in the middle layer} $f_z$ is a component perpendicular to the magnetization). The change of character of the correlations is due to the fact that the middle layer $F_2$ has a magnetization rotated by 90$^\circ$ with respect to the outer layers. 

The analysis of the Gor'kov functions shows the effect discussed in the previous section and establishes the important different behavior of the multilayer (\dDW) as compared to the continuous rotation of the magnetization of a domain wall such as in an XS (\cDW\ systems). 
The essential point made by Houzet and Buzdin is shown explicitly with the plot of the Gor'kov functions in Fig.~\ref{S3FSgorkov} which is that the $m=0$ component play no role in the center layer if $h_{F_2}$ is strong enough.  The multilayer of Ref.~\onlinecite{houzetPRB07} was chosen to suppress the $m=0$ components {in the middle layer.} 
In Fig.~\ref{S3FSgorkov} the singlet (solid blue line) and $m=0$ triplet (dashed red line) components indeed decay on a length scale $\xi_F$.

{It is important to realize that the Gor'kov functions plotted in Fig.~\ref{S3FSgorkov} are obtained from singlet Cooper pairs leaking from the {\it left S only} (a similar mirrored figure would result from Cooper pairs leaking from the right S). Thus, the $m=0$ components found near the F$_2$F$_3$ interface are the result of the cascade effect in the spin valve structure where $m=0$ components are regenerated by the rotation of the magnetization at that interface.} This resurgence is the only signature of the cascade effect in the \dDW\ as can be seen by comparing Fig.~\ref{S3FSgorkov} and Figs.~\ref{varCoGorkov},\ref{Gorkovf0XS}.  {The resurgence of the $m=0$ correlations deep in the multilayer of misaligned homogeneous Fs can be brought to light by measuring the current through} a pentalayer spin valve {as} detailed in Ref.~\onlinecite{bakerEPL14}.

An interesting consequence of this difference in the behavior of $m=0$ components between \dDW\ and \cDW\ was pointed out in Ref.~\onlinecite{bakerNJP14}, namely the occurrence of a $0-\pi$ transition of a new kind, and is discussed in the next section.

\section{Experimental Consequences of Pair Correlation Mixing: The Josephson Current}\label{sec: jcurrent}

The previous sections presented an analysis of the Gor'kov functions in exchange springs, helical structures (\cDW\ class) and misaligned homogeneous multilayers (\dDW\ class). We showed that there are important differences in the diffusion of pair correlations through continuous and discrete rotating magnetizations. In this section and the next we discuss how these differences affect the Josephson critical current and propose a general classification of $0-\pi$ transitions (Josephson current reversals; see Table \ref{tab:0pi}).

To calculate the Josephson critical current in first harmonic ($\propto \sin\varphi$), we use Eq.~\eqref{IcRNtotdecomp}, which involves {sums over}  all Matsubara frequencies. {In previous sections we showed pair correlations for $\omega_{n=0}$ only. The other frequencies display same features albeit slightly shifted or reduced in magnitude and are taken into account in the following.

\subsection{General properties of \cDW\ and \dDW\ for the Josephson current}\label{ss:generalJosephson}

  {The behavior of the Gork'ov functions presented in the previous section leads to four observations on how they impact the Josephson current.} 

First, the difference between discrete (\dDW) and continuously (\cDW) rotating magnetic structures is in the $m=0$ components.
{Although both types of magnetic structures generate $m\neq 0$ components that dominate correlations well beyond one coherence length $\xi_F$ and contribute to the Josephson current, the $m = 0$ correlations are only generated throughout in a \cDW; the \dDW\ generally studied is designed to isolate either $m=0$ or $m\neq 0$ components.\cite{houzetPRB07,bakerEPL14} In a \cDW\ the $m=0$ components affect the Josephson current and in particular the $0-\pi$ transition. \cite{bakerNJP14} This is a consequence of the cascade effect.\cite{bakerEPL14} }

Second, the presence of $m\neq 0$ pair correlations in a proximity system with a singlet pair superconductor is largest if there is a homogeneous magnetization region of thickness $\sim \xi_F$ near the SF interface.\cite{houzetPRB07,khairePRL10,buzdinPRB11} This thin homogeneous region allows the $m=0$ triplet component to develop to its maximal value (the middle of a 'hump' of the Gor'kov function) before being transformed into the $m\neq 0$ components by a subsequent rotation of the magnetization. This is the reason for choosing $F_{1,3}$ with thicknesses $d_{F_1}\sim d_{F_3}\sim \xi_F$ in the spin valve structure of Fig.~\ref{S3FSgorkov}. The same feature appears in \cDW\ such as the XS where flat regions with $\phi(x)$ nearly constant are found near the edges of Fig.~\ref{XSDWs}.\cite{bakerAIP16}} 
Helixes described by Eq.~\eqref{hhelical} can simulate such flat SF edges {as well, but the width of this region is determined by the value of $Q$ and is thus of order $\xi_F$ only for specific systems.\cite{bakerAIP16} } 

Third, the XS is a bilayer, yet we observe a Josephson current in the first harmonic. As pointed out in Ref.~\onlinecite{bakerNJP14}, our result refines the statements made in Ref.~\onlinecite{richardPRL13} (and Ref.~\onlinecite{trifunovicPRL11}) about the existence of a Josephson current through a magnetic bilayer. In the XS, the two layers are coupled magnetically, which results in the formation of a domain wall rather than two misaligned homogeneous layers. The conclusions of Ref.~\onlinecite{richardPRL13} applies to the latter, not the former.  The XS is also not equivalent to the Nb/Ho/Co/Ho/Nb case of Ref.~\onlinecite{robinsonS10} based on the behavior of the Gor'kov functions and the profiles in Fig.~\ref{XSDWs}. In this latter {system}, the magnetization of Ho rotates over a very short distance, of the order of a few nanometers.  As stated in the previous section, this type of helical layer is thus more akin to a spin active interfaces than a multilayer of misaligned homogeneous Fs.

Fourth, the results on the XS show that a Josephson current can be observed in the presence of an {\it asymmetric} magnetic structure. There is no physical reason imposing the symmetric choice, as long as the magnetic structure allows for the generation of the $m\neq 0$ components at both SF and FS interfaces.

\subsection{Classification of $0-\pi$ transitions of the Josephson current}\label{ss:classificationcurrent}

{In Table \ref{tab:0pi}  we classify $0-\pi$ transitions of the Josephson current according to the symmetry of pair correlations involved in the generation of the current.}  In the experimental observation of this effect, one measures the Josephson current as a function of some external parameter (thickness of the magnetic layer, twist of the magnetization, temperature, etc.). Keeping constant all but that one parameter, it is observed that the current changes sign as the parameter value is increased. The experiment does not reveal the reason for the change in current direction. {This insight is provided by} the Gor'kov functions {and} leads us to distinguish three types of $0-\pi$ transitions (see Table \ref{tab:0pi}): one involving only $m=0$ {correlations}, one involving only $m\neq 0$ components and one involving both $m=0$ and $m\neq 0$ {correlations}.
 \begingroup
 \squeezetable
 \begin{table*}
\caption{\label{tab:0pi} 
Classification of $0-\pi$ transitions of the Josephson current in the first harmonic, according to the pair correlations symmetries involved (column 1). The second column indicates what physical quantities can be tuned experimentally to observe the transition. Column 3 and 4 indicate the contribution of singlet and triplet pair correlations to the current. Column 5 lists the class of magnetic systems studied in this paper where the effect was or can be observed. $d_F$ is the thickness of the F, $T$ the temperature, $\Delta\phi$ the angle difference between the magnetization direction on either side of the magnetic layer.}
\begin{ruledtabular}
\begin{tabular}{lllll}
Determinant & Variable& Singlet $(I_{c,0})$ & Triplets $(I_{c,t})$ & System \\
correlations & parameter & & \\ \hline
Singlet\footnote{Buzdin--Bulaevskii--Panyukov in Ref.~\onlinecite{buzdinJETP82}.} & $d_F$ or  $T$ & $I_{c,0}(d_F)$ changes sign with $d_F$ & $I_{c,t}=0$ & SFS (\dDW) \\
Triplet \footnote{Houzet--Buzdin in Ref.~\onlinecite{houzetPRB07}.} & $\Delta\phi$ & $I_{c,0}=0$ & $I_{c,t} (\Delta\phi)$ changes sign with $\Delta\phi$\\
Singlet-triplet \footnote{Ref.~\onlinecite{bakerNJP14} and mixed with type $a$ in Ref.~\onlinecite{bergeretPRB01}. These two references describe different transitions (see text).} & $\Delta\phi$ or $T$ & $I_{c,0} (\Delta\phi)>0$ (Fig.14a) & $I_{c,t} (\Delta\phi)$ increases with $\Delta\phi$ and  &  XS, helix (\cDW)\\
& $I_{c,0}$ opposite to $I_{c,t}$ &   (determined by $d_F$ and $\varphi$) & has definite sign (negative). & 
\end{tabular}
\end{ruledtabular}
\end{table*}
\endgroup
As we {now} discuss, the physical mechanism behind an experimentally observed $0-\pi$ transition is quite different depending on the structure of the magnetic multilayer embedded into the Josephson junction.

The $0-\pi$ transition phenomenon has first been predicted by Buzdin, Bulaevskii and Panyukov in Ref.~\onlinecite{buzdinJETP82} (see also Ref.~\onlinecite{buzdinJETP91}) to occur in a junction where a homogeneous F is sandwiched between two singlet pair superconductors S (Fig.~\ref{fig: jcdens}b). They pointed out that the oscillation of the $m=0$ (singlet and triplet) pair correlations in F may lead to a reversal in direction of the Josephson current as one increases the thickness $d_F$ of the F under otherwise identical experimental conditions. The effect was later observed in Refs.~\onlinecite{ryazanovPRL01,kontosPRL02,oboznovPRL06,pianoEPJ07}. The reason for this transition is the change of relative sign between the left and the right contributions of the $m=0$ Gor'kov functions.  {As one increases the thickness} an extra node {appears} in the Gor'kov functions  {that} causes terms like $f_{-n}^*\partial_xf_n$ to change sign in Eq.~\eqref{IcRNtot} (see Fig.~\ref{fig: jcdens}a). This results in {the familiar} $j_c \propto \cos(d_F/\xi_F)$ dependence.\cite{bergeretRMP05,buzdinRMP05} The relevant lengthscale that determines the physics of the Buzdin-Bulaevskii-Panyukov $0-\pi$ transition seen in Fig.~\ref{fig: jcdens}b is $\xi_F$, which is typically of the order of a few nanometers. {This transition solely involves}  $m=0$ singlets and triplets {pair correlations}. 

A different type of $0-\pi$ transition of the Josephson current has been predicted by Houzet and Buzdin in Ref.~\onlinecite{houzetPRB07}. The generic magnetic structure for that novel type of $0-\pi$ transition is the spin valve structure of Fig.~\ref{spinvalvediagram}, which we remind belongs to the \dDW\ class.  The corresponding Gor'kov functions are shown in Fig.~\ref{S3FSgorkov}. This structure was chosen with a thickness of the middle layer F$_2$ large enough to supress the $m=0$ components completely ($d_{F_2} \gg \xi_F$), which implies that no Buzdin-Bulaevskii-Panyukov  $0-\pi$ transition of the current will be observed. Instead, only $m\neq 0$ components are long range enough and dominate pair correlations across the layer F$_2$.  Starting with the configuration of Figs.~\ref{spinvalvediagram}, \ref{S3FSgorkov} where the magnetization is oriented along the $\hat{\mathbf{z}}$ axis in F$_{1,3}$ we rotate for example F$_3$.  The $0-\pi$ transition occurs when the components of the magnetization along $\hat{\mathbf{z}}$ in F$_1$ and F$_3$ are opposite in sign; since F$_1$ has magnetization along $\phi_1 = \pi$ the transition occurs when $\phi_3 \leq \pi/2$.\cite{houzetPRB07} We emphasize that in this scenario the $m=0$ {\it plays no role} since the structure was designed to suppress these components. The Houzet-Buzdin $0-\pi$ transition is thus conceptually different from the Buzdin-Bulaevskii-Panyukov current reversal. 

Finally, a third, distinct mechanism for the $0-\pi$ transition was {proposed} in Ref.~\onlinecite{bakerNJP14} and involves {\it both} $m=0$ and $m\neq 0$ components. Because the reversal of the current is due to the competition of singlet and triplet contributions it is termed a {\it singlet-triplet} $0-\pi$ transition.  {This class of $0-\pi$ transition is here observed in \cDW s.} The {defining} features {of this transition} are not readily seen in the figures of the previous section where only the magnitudes $|f_j|$ ($j=0,y,z$) are displayed. Rather, we {need to} consider the {different contributions to the} expression of the current, Eq.~\eqref{IcRNtotdecomp}, as discussed in more detail below. In absence of a domain wall (homogeneous case) the critical current {is due to $m=0$ components, $I_c(\Delta\phi=0)=I_{c,0}$. The direction (sign) of the current} is determined by the phase difference $\varphi$ between the two superconductors and the thickness of the magnetic layer.\cite{buzdinJETP82,buzdinRMP05} The inhomogeneous case is different in that also the {\it relative} sign of the components $I_{c,0}$ and $I_{c,\alpha}$ (with $\alpha=x,y,z$ or $\alpha = \perp,\parallel$) matters for the direction of the Josephson critical current and the observation of a $0-\pi$ transition.

The key ingredient for making a singlet-triplet $0-\pi$ transition is to {choose the thickness of the magnetic structure so that the singlet contribution to the current is opposite to that of the $m\neq 0$ contributions (the sign of each contribution is a matter of convention). In the examples discussed below, the singlet contribution $I_{c,0}$ will be chosen positive.} ({\it i.e.}, the untwisted XS is in the middle of a $0-${phase} dome of the oscillatory $m=0$ component -- see for example the solid blue line in Fig.~\ref{S3FSgorkov}).  {In this situation,} the sign of the $m\neq0$ contribution to the current, $I_{c,t}$, is always negative. The $m=0$ and $m\neq 0$ contributions to the current vary at different rates when increasing the twist $\Delta \phi$ of the magnetization; they compete. In certain instances the current contribution from the $m\neq 0$ correlations overcomes that of the singlet, leading to a change of direction of the total current. This case is discussed in more detail next.

\subsection{The singlet-triplet $0-\pi$ transition in \cDW s}\label{ss:singlet-triplet}

We examine the singlet-triplet $0-\pi$ transition of the Josephson current in the \cDW s to elucidate the conditions under which it can be observed. The different pair-correlation contributions to the current are presented in Fig.~\ref{signedjcvarCoNi} for the Ni- and Co-XS and in Fig.~\ref{signedjchelix} for different helixes.

We observe first that both the XS and helical domain walls can exhibit {the singlet-triplet} $0-\pi$ transition, though the effect {is easier to realize} experimentally in the XS with current methods since it is tunable. 

In the XS systems {of the previous section}  {we note that only the} Ni-XS displays the $0-\pi$ transition. Due to its strong magnetization the Co-XS has a {vanishingly small singlet contribution in the homogeneous case; the small  contribution seen in Fig.~\ref{signedjcvarCoNi} is only generated through the cascade effect for a sufficient twist of the magnetization}.  Hence, the Co-XS does not allow for a {singlet-triplet} $0-\pi$ transition.  Note that it is also possible to eliminate the $0-\pi$ transition in Ni-XS by simply changing the thickness of the XS, which changes the sign of $I_{c,0}$ and removes the competition between singlet and triplet contributions.\cite{bakerNJP14}

\begin{figure}
\begin{center}
\includegraphics[width= \columnwidth]{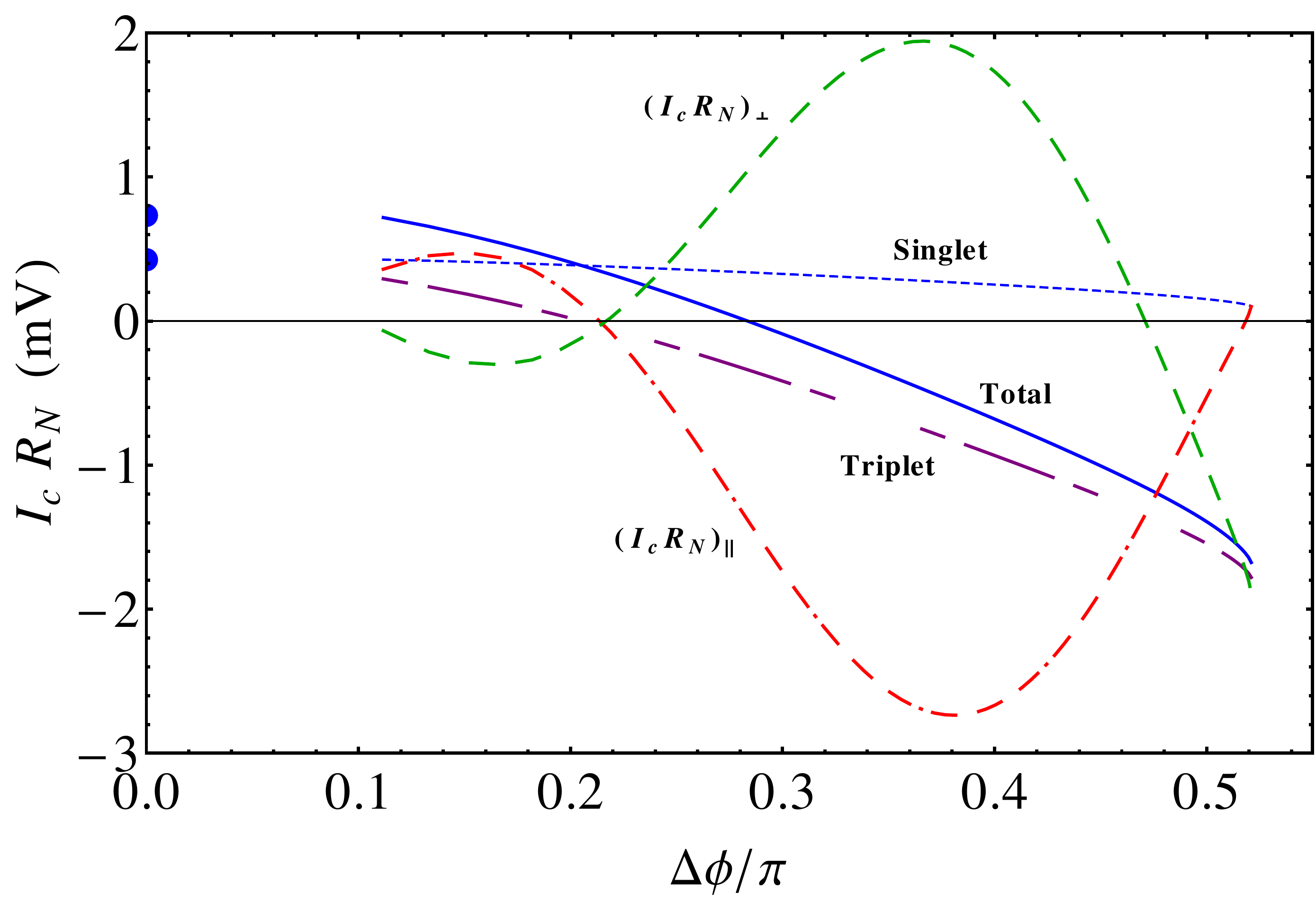}
\includegraphics[width= \columnwidth]{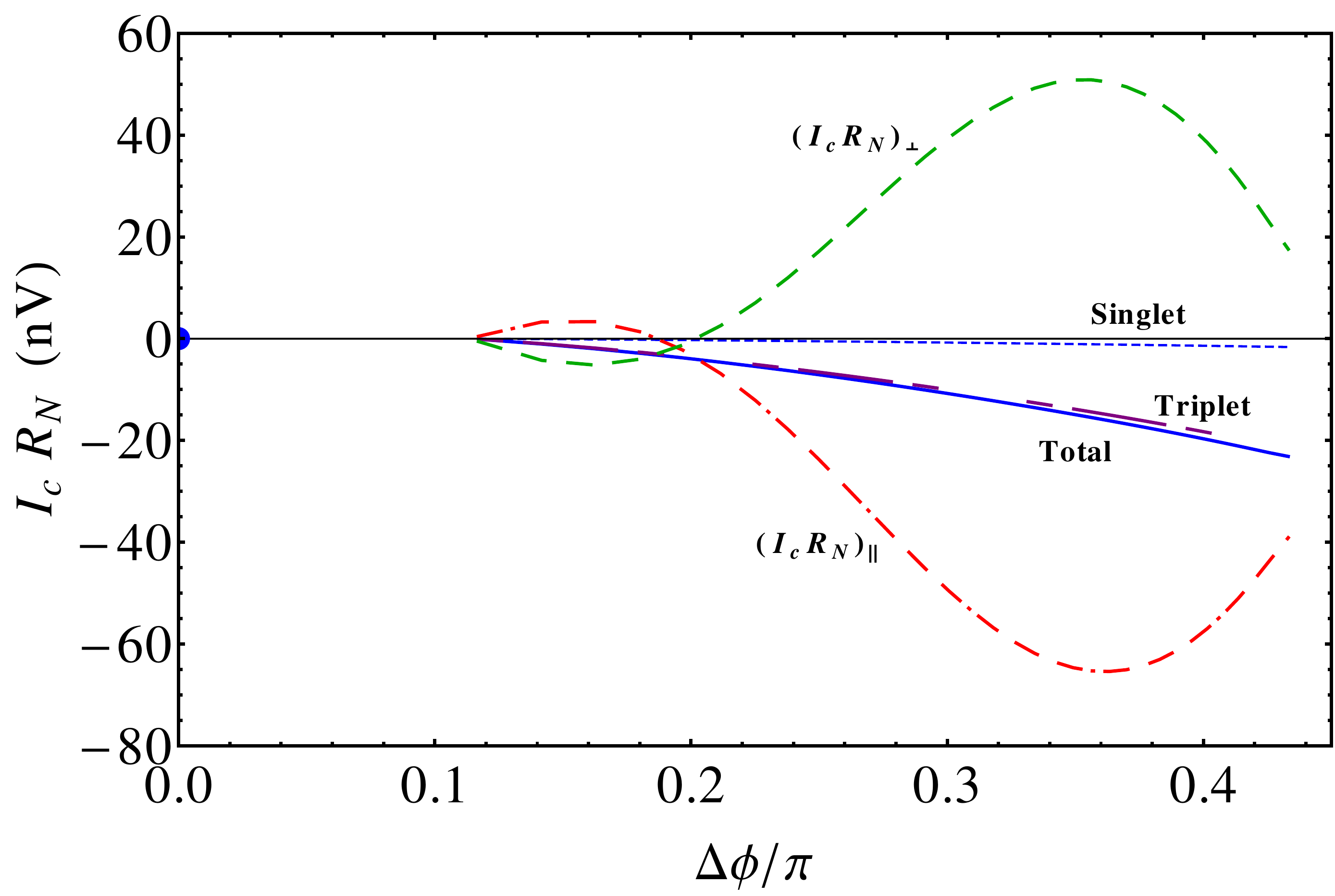}
\end{center}
\caption{\label{signedjcvarCoNi} (color online) {\it Signed} contributions $I_{c,0}$ (singlet) and $I_{c,t} = I_{c,\parallel} + I_{c,\perp}$ (triplet) to the total Josephson current $I_c(\Delta\phi)$. top: Ni-XS, bottom: Co-XS. Only the total current $I_c(\Delta\phi) = I_{c,0}+I_{c,t}$ is measurable. The sign and weight of the different contributions to the current obtained from the Gor'kov functions allow understanding why Ni-XS displays a singlet-triplet $0-\pi$ transition and why Co-XS does not.}
\end{figure}
\begin{figure}
\begin{center}
\includegraphics[width= \columnwidth]{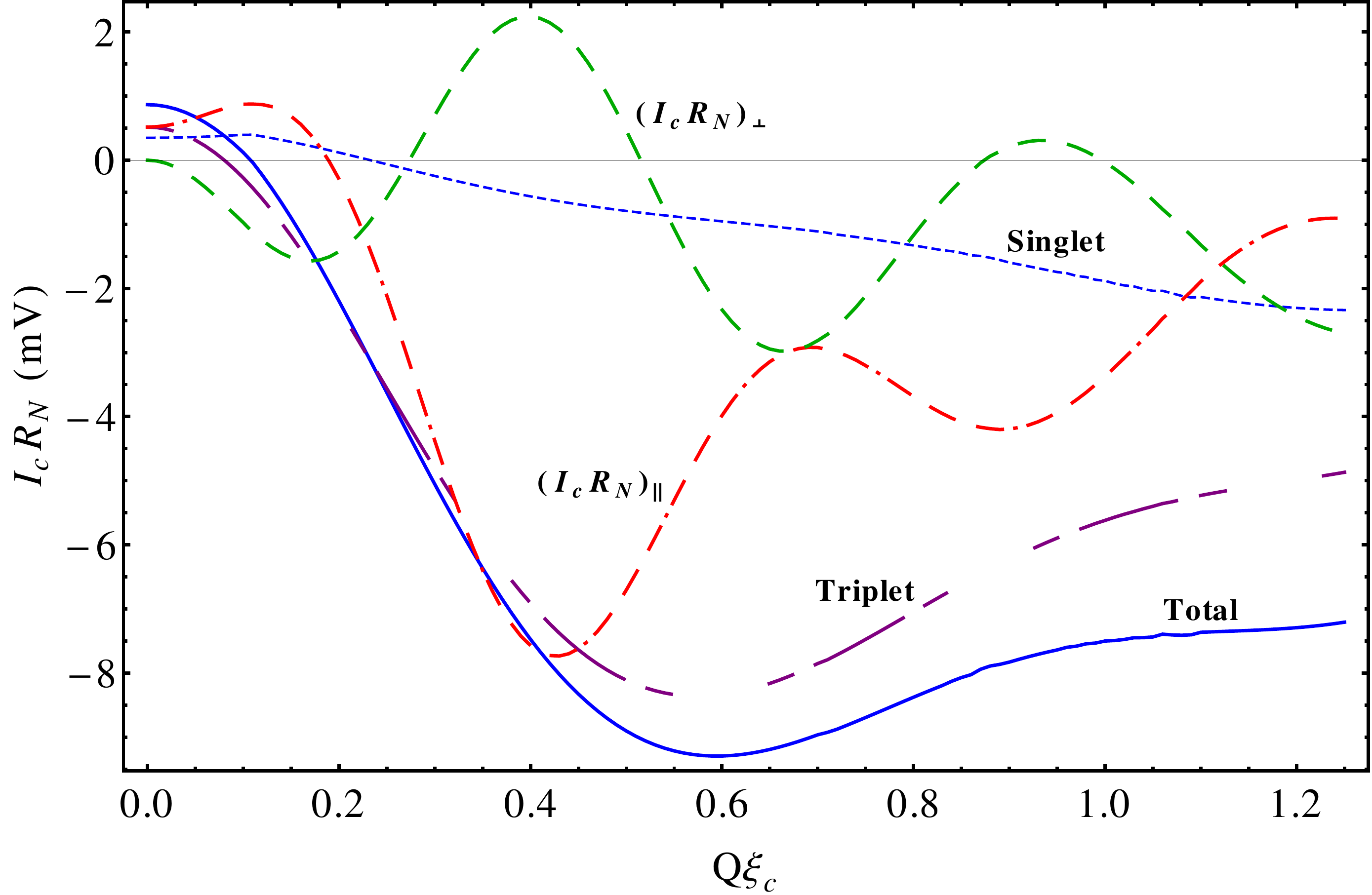}
\end{center}
\caption{\label{signedjchelix} (color online) Signed singlet and triplet contributions to the total current for a helix of varying $Q$.  The same linestyle applies as in Fig.~\ref{signedjcvarCoNi}. Note that changing $Q$ on the horizontal axis is equivalent to changing the magnetic material in the system. The current saturates to $\approx-203$mV for $Q\rightarrow\infty$, which is the value of the current through a normal metal ($h=0$, singlet components only), implying the triplet correlations decay to zero. Parameters used are: $d_f=7\xi_c$, $h=8\pi T_c$, and $T=0.4T_c$.}
\end{figure}

In the instance of the Ni-XS (see Ref.~\onlinecite{bakerNJP14}), we picked a thickness of the magnetic system such that the singlet contribution to the current is positive.\cite{buzdinJETP82,buzdinJETP91} It is seen in Fig.~\ref{signedjcvarCoNi} that this contribution remains a positive, weakly varying function of the twist, while the $m\neq0$ components also contribute negatively to the current and grow with $\Delta\phi$ at a higher rate. Hence, the twist of the magnetization can be increased until the current contribution of the $m\neq0$ components counter-balances the singlet contribution at which point the current vanishes and produces the node of the $0-\pi$ transition. As one continues increasing the twist, the current changes sign. This explains how the $0-\pi$ transition of Ref.~\onlinecite{bakerNJP14} is different from other transitions presented in the literature and grouped in the two first rows of table \ref{tab:0pi}. We note that the relative weight of the competing contributions is essential and differentiates the strong ferromagnet Co-XS from the weak ferromagnet Ni-XS.

Figure~\ref{signedjchelix} displays the contributions of pair correlations to the current for helixes as a function of $Q$. In contrast to the XS the singlet contribution changes sign with the tuning of the inhomogeneity through the change of $Q\xi_c$.  On the other hand, $I_{c,\perp}$ and $I_{c,\parallel}$ oscillate with $Q$ and leads to a feature that is common with the XS case: the oscillations of $I_{c,\perp}$ and $I_{c,\parallel}$ are out of phase, leading to a much smoother current $I_{c,t}$. This reminds of the simplest SFS case of Fig.~\ref{fig: jcdens}a. Finally, as noted in Ref.~\onlinecite{bergeretPRB01} and seen in Fig.~\ref{signedjchelix} for $Q\xi_c \lesssim 0.2$, at fixed low temperature the critical current undergoes a $0-\pi$ transition as one increases the value of $Q\xi_c$. Note that this transition is more difficult to realize experimentally since $Q$ is not tunable with an external perturbation.

The progressive twisting of the helix in Fig.~\ref{signedjchelix} conveys a novel experimental result.  The interplay between $f_\perp$ and $f_\parallel$ causes a minimum in the value of the signed Josephson current $I_c$, at $Q\xi_c\approx0.6$.  This is a local minimum as a function of $Q$ since a steady increase of the singlet component is expected as $Q\rightarrow\infty$, which approaches an antiferromagnet and is modeled effectively as a normal metal, $h=0$, with a current entirely due to the singlet contributions.

As previously remarked on for example in Refs.~\onlinecite{buzdinJETP82,bergeretPRB01,ryazanovPRL01,oboznovPRL06,houzetPRB07,bakerNJP14}, the $0-\pi$ transition can be induced by varying temperature keeping all other parameters fixed. We note that there is a competing effect between temperature and twist of the magnetization. An increase in temperature reduces the superconducting condensate. Concomitantly, an increased twist leads to stronger triplet correlations in the magnetic system. In the XS the nodes of the $0-\pi$ transition in $I_c(T)$ are shifted to lower temperatures as one increases the twist $\Delta\phi/\pi$ of the magnetization.\cite{bakerNJP14}

Bergeret and co-workers calculated in Ref.~\onlinecite{bergeretPRB01} the critical current as a function of temperature through a helical structure, choosing a thickness such that for the homogeneous case the current is close to the $0-\pi$ Buzdin-Bulaevskii-Panyukov transition. They considered a weak rotation of the magnetization, $Q\xi_c \lesssim 0.2$ and observe a $0-\pi$ transition.  The results of Ref.~\onlinecite{bergeretPRB01} are qualitatively different from the singlet-triplet transition discussed here and in Ref.~\onlinecite{bakerNJP14}.  To see this, note first that Ref.~\onlinecite{bergeretPRB01} has tuned the thickness of their F layer so it is close to the Buzdin-Bulaevskii-Panyukov $0-\pi$ transition, which could be induced by varying $T$. If the thickness of the homogeneous F is chosen away from that particular case, then the system does not undergo a $0-\pi$ transition with $T$.  With this choice of F thickness, the $m\neq0$ components can perturb the Buzdin-Bulaevskii-Panyukov type transition.  This special situation is evidence for the singlet-triplet $0-\pi$ transition since it shows that the $m=0$ and $m\neq 0$ correlations can affect one another. On the other hand, it does not show that the triplet components can overcome the singlet contribution on their own.  That is demonstrated in the XS layers from Ref.~\onlinecite{bakerNJP14}.

One could be tempted to state that the singlet-triplet $0-\pi$ transition is a particular limit of the Houzet-Buzdin transition when the magnetization in the sample is weak. This is, however, not the case. {Reducing} the magnetization in the central F$_2$ layer of the spin valve structure leads to an increase of the $m=0$ components, thereby coming close in magnitude to the triplet component near the interfaces. However, since the layer has homogeneous magnetization these $m=0$ correlations still decay on the lengthscale $\xi_F$ and oscillate, changing sign in the layer.  This contrasts with the situation encountered in the singlet-triplet transition where a sustained generation of same sign $m=0$ correlations is obtained by the continuously rotating magnetization. We thus emphasize that the singlet-triplet $0-\pi$ transition is not a simple sum of the Buzdin-Bulaevskii-Panyukov  and Houzet-Buzdin effects. It relies on a more subtle balance between $m=0$ and $m\neq 0$ pair correlations and is found in a continuously rotating magnetization while the two other transitions are found in a discrete rotating magnetization.

We note that there are situations where a singlet-triplet transition may be observable in dDW heterostructures. These are more complicated than those studied here and are out of the scope of this paper.

We also point out that the magnitude of the calculated currents in the singlet-triplet $0-\pi$ transition are not small compared to other $0-\pi$ transitions of table \ref{tab:0pi}. For example, the current amplitude through the Ni-XS of Ref.~\onlinecite{bakerNJP14} is of the {\it same} order as the currents calculated in Ref.~\onlinecite{houzetPRB07} when converted to the same units.

Finally, the results of this and the previous section confirm the statement made earlier that the distinctions between short and long ranged components are most meaningful in discrete magnetization rotation configurations. The distinction is less useful for the class of continuously rotating magnetizations.

\subsection{Possible scenarios for observing Josephson currents through an XS}\label{sec: expt}

In this section we make a few general comments on the properties of materials and the structure of the XS that may serve as suggestions for the experimental study of magnetic Josephson junctions with an XS.

\paragraph{Experimental realization of S/XS/S junctions.} 
Ref.~\onlinecite{bakerNJP14} proposes several practical ways to implement the theoretically proposed S/XS/S structure and how to observe the studied effects, in particular the singlet-triplet $0-\pi$ transition of the Josephson current. {The work done by Gu {\it et.~al.} in Ref.~\onlinecite{guPRB10} should be extended to generate a Josephson junction and an improved clean magnetic structure.} Ref.~\onlinecite{bakerEPL14} also suggests an experiment where the $m=0$ components can be shown to exist deep in the magnetic material and matter for the Josephson current.  Next to these suggestions we propose here another possible candidate to measure these effects: a robust BCS superconductor {such as} MgB$_2$\cite{burnellAPL01,uedaAPL05} and a highly anisotropic exchange spring GdFe/TbFe \cite{manginPB00} together with a thin metallic film to tune the interface coupling (see below).  Though this exchange spring has an extra anisotropy axis and will not be described by the model in the formulation presented here, the principles of our work still apply.
This material may also allow one to place two exchange springs sandwiching a normal metal region or, if feasible, a ferromagnet to show that the tunneling occurs over very long lengths.\cite{porterPRB15}

To observe the singlet-triplet Josephson current reversal one needs an XS with high anisotropy ratio between the hard and soft F, small width but large enough for the XS to generate a domain wall, and relatively weak magnetization strength. The thinner the layers, the higher anisotropy ratio $K_h/K_s$ is needed to allow a fuller domain wall (and more dramatic twist) to appear with a smaller applied magnetic field.  Altering these parameters may help or harm an experimental investigation of the effects we consider.

In this and previous work, we consider an XS made of hard and soft ferromagnetic layers of fixed, constant thickness.  It would be of interest to extend the study to other types of XSs. For example, one could imagine that both layers have wedge form, keeping the total thickness of the bilayer constant; for example, the hard (soft) F would have maximal (minimal) thickness at top of the bilayer and vice-versa at the bottom. One could conceive an XS where the hard layer has constant thickness while the soft layer is a wedge. Or an XS where both layers have wedge form with minimal thickness on the same end. The study of these alternative systems goes beyond the present work, but they are expected to display a richer inhomogeneity of the magnetization and new features of the pair correlations and Josephson current.

\paragraph{Magnetization strength.} The magnetization strength $h$ plays an important role since for example $\xi_F \propto h^{-1/2}$ in the diffusive regime. {As stated earlier, $\xi_F$ is the approximate width of homogeneous magnetic material required at the SF interface} to obtain maximal $m=0$ triplet correlations.\cite{buzdinPRB11}  Effectively, one needs an edge with weak curvature of $\phi(x)$ so as to ensure that the singlets have ample opportunity to transition to the $m\neq0$ component through the $m=0$ triplet.  
The beauty of the exchange spring magnetic domain wall is that it naturally provides for a region at the interface where the magnetization is weakly rotating (Fig.~\ref{DWsdiag}){, and is a result of the magnetic boundary conditions that the XS satisfies.\cite{bakerAIP16}  The width of this nearly homogeneous ferromagnetic region is tunable through an appropriate choice of the XS's magnetic anisotropy ratio.}

\paragraph{Interface between hard and soft Fs.} The boundary conditions {for the Gor'kov functions at the interface between the two magnetic films of the XS described earlier are common in the literature:} perfect transparency with equal values and derivatives of the functions at the interface.  
We discuss here instead the {\it magnetic} coupling at the interface between the hard and the soft Fs. This interaction is an essential component and notable distinction between the XS and other hybrid systems.\cite{bakerNJP14,bakerAIP16} For example, this coupling leads to the domain wall profile instead of simple misalignment of homogeneous Fs and to the presence of a Josephson current in first harmonic, even in bilayer structures. It is known that the interface magnetic coupling can be tuned by using the properties of the RKKY interaction; inserting a thin metallic layer between the Fs allows to tune the interaction and even to choose between ferromagnetic and antiferromagnetic coupling between neighbor layers.\cite{parkinPRL91} This was used in recent experiments.\cite{gingrichNP16} For the XS the tuning of the magnetic coupling between hard and soft Fs leads to a discontinuity of the rotating magnetization at the interface between hard and soft Fs. 

Experimentally, the exchange interaction constant is very nearly equal for all ferromagnets composing an XS.\cite{bozorth03}  Hence, throughout this work we assumed that the exchange interaction is the same in the hard and soft F.  We point out that any other choice would induce a kink in the domain wall profile at the interface between ferromagnets (Fig.~\ref{XSDWs}).   If the exchange interaction constants do differ it would aid the appearance of $m\neq0$ components by increasing the curvature of the domain wall at the interface. The interface magnetic interaction and the exchange interaction within each material are {knobs available to experimentalists and material scientists} to shape the domain wall in a variety of ways by inducing discontinuities in $h(x)$ and $dh(x)/dx$.

The features enumerated in this section are expected to lead to further rich physics by allowing the tuning of the magnetic profile from smooth, continuous (partial) domain wall to a misaligned homogeneous bilayer {with a variety of magnetic configurations in between}.  Changing the interlayer coupling with a metallic layer of different thicknesses is one way available to achieve that goal.

\section{Conclusion}\label{sec:conclusions}

We provided a comparative study of singlet and triplet pair correlations in magnetic proximity systems and Josephson junctions with a variety of magnetic configurations of the Bloch-type. We were led to two major conclusions. 

The first conclusion is based on the analysis of the Gor'kov functions that represent the pair correlations.  We are led to
distinguish two classes of magnetic systems: discrete domain walls (dDW) which are composed of a stack of layers with homogeneous but misaligned magnetization, and continuous domain wall (cDW) that display a continuous rotation of the magnetization. Spin valves are examples pertaining to the first class and were the most widely studied in the literature. Examples of the second, cDW class are XSs and helixes (such as Ho).

We showed that pair correlations are different in the dDW and cDW classes. While in the dDW $m=0$ correlations (the singlet and triplet usually termed short range components) are only generated at the interfaces between misaligned Fs, and decay over the length $\xi_F$ away of these interfaces, the continuous rotation of the magnetization in cDW implies a continuous generation of {\it all} components ($m=0,\pm 1$) throughout the system due to the cascade effect.\cite{bakerEPL14} In particular, singlet pair correlations can be found deep in the magnetic material of a cDW. Further, the Gor'kov function $f_0$ of singlet pair correlations is affected by the magnetization profile, via a cascade effect from $m\neq0$ components, even though that function is a scalar and should only be affected by the amplitude of the magnetization.

The second main result is to propose a classification into three types of $0-\pi$ transitions of the Josephson critical current (current reversal upon variation of one parameter of the system) and is summarized in table \ref{tab:0pi}. The classification is made according to the pair correlations symmetries involved in the Josephson current and its reversal. The first transition proposed by Buzdin, Bulaevskii and Panyukov\cite{buzdinJETP82} involves only $m=0$ pair correlations. The second transition discussed by Houzet and Buzdin\cite{houzetPRB07} involves only $m\neq 0$ pair correlations. Finally, the third type of $0-\pi$ transition proposed in Ref.~\onlinecite{bakerNJP14} involves a competition of $m=0$ and $m\neq 0$ correlations. The analysis of this paper clearly shows that while the two first types of transitions can be found in a dDW, the latter transition is of a different kind that can be found in cDWs.  The XS is a system of choice in observing that type of $0-\pi$ transition.

The paper focused on the properties of the XS proposed in Ref.~\onlinecite{bakerNJP14} to generate a Josephson junction with tunable and reversible current. The XS is an attractive component for superconducting spintronics applications as it allows for a tunable magnetic inhomogeneity in form of a partial to full Bloch domain wall in the system by applying a small external magnetic field that does not affect the superconducting properties of the system appreciably.
The parameters and thicknesses of the XS necessary to observe the correlations and new $0-\pi$ transition in wide junctions in the diffusive regime are not arbitrary and our theoretical study of XSs with different parameters (such as the strength of the magnetization in the bilayer) invites for an experimental realization of these hybrid structures. Our study also leads to several experimental suggestions that we encourage to test.

The work shows that misaligned homogeneous Fs, helical structures and XSs are clearly distinct in the way superconducting pair correlations transform and spread into the magnetic material. The exchange spring provides a unique experimental tool to probe the rich physics that magnetic Josephson junctions with inhomogeneous magnetization can display.

\section{Acknowledgements}
We gratefully acknowledge funding provided by the National Science Foundation (DMR-1309341). A.B. thanks F. Guinea and the ICMM for hospitality where this work has been started. T.E.B. gratefully acknowledges the Pat Beckman Memorial Scholarship from the Orange County Chapter of the Achievement Rewards for College Scientists Foundation.

\bibliography{SXSS,notes}

\begin{thebibliography}{88}
\bibitem [{jcl()}]{jcleftrightnote}%
  \BibitemOpen
  \href@noop {} {\ }\bibinfo {note} {Notice that if we place
  Eq.~\eqref{leftright} into Eq.~\eqref{IcRNtot}, we obtain $\frac{2e R_N}{ \pi
  T} I_c(x) = \sin\varphi\sum_{n = -\infty}^\infty \sum_{\alpha=0,y,z}\left(
  f_{-n,\alpha,L}^\star\frac{\partial f_{n,\alpha,R}}{\partial
  x}-f_{-n,\alpha,R}^\star\frac{\partial f_{n,\alpha,L}}{\partial
  x}\right)$.}\BibitemShut {Stop}%
\bibitem [{hel()}]{helixanglenote}%
  \BibitemOpen
  \href@noop {} {\ }\bibinfo {note} {The cone angle named $\alpha$ in
  Refs.~\onlinecite{fritschJPC14,fritschNJP14} ($\theta$ in
  \onlinecite{wuPRL12,wuPRB12}) is set to $\pi/2$ while their angle $\beta$
  ($\varphi y/2$ in \onlinecite{wuPRL12,wuPRB12}) is related to our angle by
  $\beta =\varphi y/2 = \pi - \phi$. Eq.~\eqref{hhelical} is also exactly as in
  Ref.~\onlinecite{bergeretPRB01} with their angle $\alpha =
  \pi-\phi$.}\BibitemShut {Stop}%
\end{thebibliography}


\end{document}